\documentclass[aps,prb,twocolumn,eqsecnum]{revtex4}

\usepackage{amsmath}
\usepackage{amssymb}
\usepackage{graphicx}

\newcommand{\ocite}{\onlinecite}

\newcommand{\iy}{\infty}

\newcommand{\pd}{\partial}

\newcommand{\dg}{\dagger}
\newcommand{\lan}{\langle}
\newcommand{\ran}{\rangle}
\newcommand{\lt}{\left}
\newcommand{\rt}{\right}

\newcommand{\f}{\frac}
\newcommand{\tf}{\tfrac}

\newcommand{\sq}{\sqrt}
\newcommand{\lbl}{\label}

\newcommand{\nn}{\nonumber}

\newcommand{\tm}{\times}

\newcommand{\eq}[1]{Eq.~(\ref{eq:#1})}
\newcommand{\eqs}[2]{Eqs.~(\ref{eq:#1}) and (\ref{eq:#2})}
\newcommand{\eqss}[3]{Eqs.~(\ref{eq:#1}), (\ref{eq:#2}), and (\ref{eq:#3})}

\newcommand{\eqn}[1]{(\ref{eq:#1})}
\newcommand{\eqsn}[2]{(\ref{eq:#1}) and (\ref{eq:#2})}

\newcommand{\tabr}[1]{Tab.~\ref{tab:#1}}

\newcommand{\secr}[1]{Sec.~\ref{sec:#1}}
\newcommand{\secsr}[2]{Secs.~\ref{sec:#1} and \ref{sec:#2}}

\newcommand{\figr}[1]{Fig.~\ref{fig:#1}}
\newcommand{\figsr}[2]{Figs.~\ref{fig:#1} and \ref{fig:#2}}

\newcommand{\spc}{\mbox{ }}

\newcommand{\beq}{\begin{equation}}
\newcommand{\eeq}{\end{equation}}
\newcommand{\beqar}{\begin{eqnarray}}
\newcommand{\eeqar}{\end{eqnarray}}
\newcommand{\beqarn}{\begin{eqnarray*}}
\newcommand{\eeqarn}{\end{eqnarray*}}
\newcommand{\ba}{\begin{array}}
\newcommand{\ea}{\end{array}}
\newcommand{\bwt}{\begin{widetext}}
\newcommand{\ewt}{\end{widetext}}

\newcommand{\sgn}{{\rm sgn}\,}

\newcommand{\LRa}{\Leftrightarrow}

\newcommand{\rarr}{\rightarrow}

\newcommand{\dx}{{\text d}}
\newcommand{\ex}{{\text e}}

\newcommand{\hx}{{\text h}}
\newcommand{\ix}{{\text i}}

\newcommand{\Dx}{{\text D}}

\newcommand{\Rx}{{\text R}}

\newcommand{\Tx}{{\text T}}
\newcommand{\Ux}{{\text U}}

\newcommand{\ch}{\hat{c}}

\newcommand{\Hh}{\hat{H}}

\newcommand{\Nh}{\hat{N}}

\newcommand{\psih}{\hat{\psi}}

\newcommand{\Fc}{\mathcal{F}}

\newcommand{\Oc}{\mathcal{O}}

\newcommand{\Sc}{\mathcal{S}}
\newcommand{\Tc}{\mathcal{T}}

\newcommand{\Gt}{\tilde{G}}

\newcommand{\gt}{\tilde{g}}

\newcommand{\altd}{\tilde{\alpha}}

\newcommand{\Det}{\tilde{\Delta}}

\newcommand{\psith}{\hat{\tilde{\psi}}}
\newcommand{\Hth}{\hat{\tilde{H}}}
\newcommand{\cth}{\hat{\tilde{c}}}

\newcommand{\al}{\alpha}

\newcommand{\de}{\delta}
\newcommand{\De}{\Delta}

\newcommand{\ka}{\varkappa}

\newcommand{\eps}{\varepsilon}
\newcommand{\e}{\epsilon}

\newcommand{\hc}{\text{h.c.}}

\begin{document}
\title{Backscattering in a helical liquid induced by Rashba spin-orbit coupling\\
and electron interactions: locality, symmetry, and cutoff aspects}

\author{Maxim Kharitonov$^{1,2}$}
\author{Florian Geissler$^1$}
\author{Bj\"orn Trauzettel$^1$}
\affiliation{$^1$Institute for Theoretical Physics and Astrophysics, University of W\"urzburg, 97074 W\"urzburg, Germany\\
$^2$Donostia International Physics Center (DIPC), Manuel de Lardizabal 5, E-20018 San Sebastian, Spain}

\date{\today}

\begin{abstract}

The combination of the time-reversal-symmetric single-particle backscattering field (commonly known as Rashba spin-orbit coupling)
and non-backscattering electron interactions is generally expected to
produce inelastic backscattering in 1D helical electron liquids at the edge of 2D topological insulators,
as theoretically predicted in a number of works.
An opposite conclusion of absent backscattering was reached in a recent work
[H.-Y. Xie {\em et al.},
Phys. Rev. Lett. {\bf 116}, 086603 (2016)]
for the ``local'' model of the backscattering field and interactions.
Motivated to resolve this potential controversy, in the present work, we study backscattering effects
employing fermionic perturbation theory and considering quite general forms
of the backscattering field and electron interactions.
We discover that backscattering effects are crucially sensitive to the {\em locality} properties
of the backscattering field and electron interactions, to the {\em symmetry} of the latter,
as well as to the presence or absence of the {\em cutoff} of the electron spectrum.
We find that backscattering is indeed absent under the following assumptions:
(i) local backscattering field; (ii.a) local or (ii.b) SU(2)-symmetric interactions;
(iii) absent cutoff of the edge-state spectrum.
However, violation of any of these conditions  leads to backscattering.
This also reconciles with the results based on the bosonization technique.
We calculate the associated backscattering current, establish its low-bias scaling behavior,
and predict a crossover between two different scaling regimes.
The main implication of our findings is that backscattering of some magnitude is inevitable in a real system,
although could be quite suppressed for nearly local backscattering field and interactions.

\end{abstract}

\maketitle

\section{Introduction\lbl{sec:intro}}

\begin{figure}
\includegraphics[width=.45\textwidth]{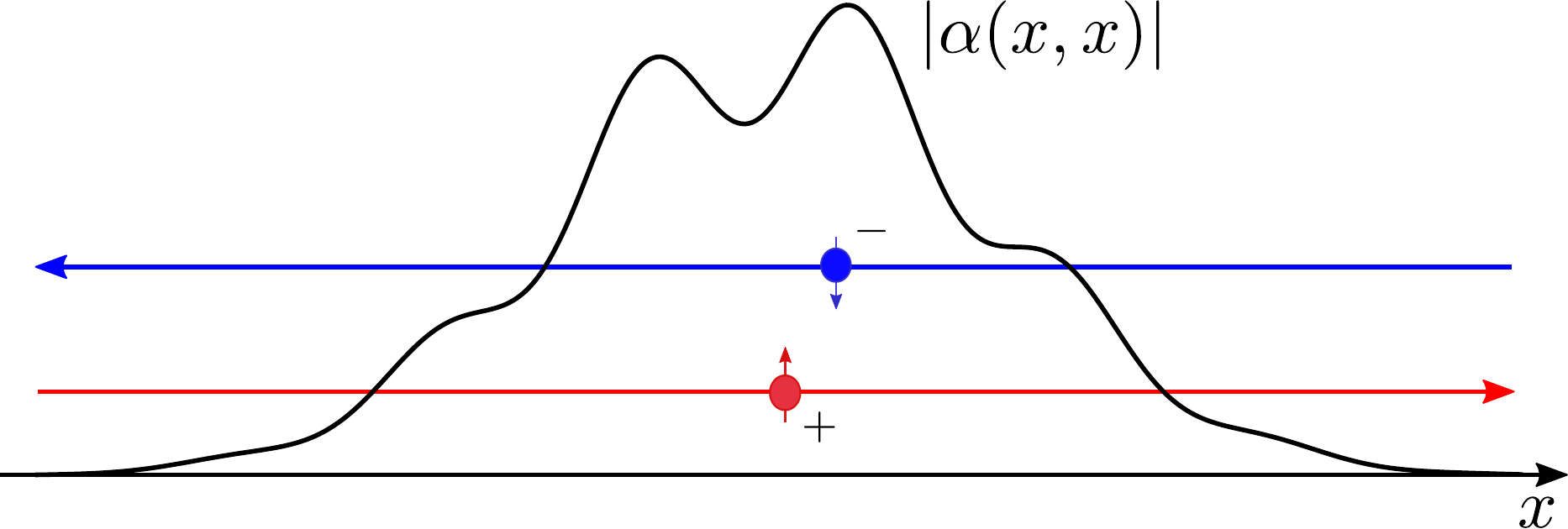}
\caption{
Schematic of the helical edge states of a 2D topological insulator and the profile $\al(x,x)$ of the BS field [\eq{HR}].
}
\lbl{fig:edge}
\end{figure}

\begin{table}
\begin{tabular}{c||c|c}
interactions$\backslash$BS field & (i) local & nonlocal \\ \hline\hline
(ii.a) local or (ii.b) SU(2)-symm. & no & yes \\\hline
nonlocal and SU(2)-asymm. & yes & yes \\
\end{tabular}
\caption{
Conditions for the presence (yes) or absence (no) of backscattering (BS),
depending on locality of the BS field and interactions
and symmetry of the latter, for absent cutoff of the electron spectrum.
If a cutoff is present, BS is present regardless of the forms of the BS field and interactions.
}
\label{tab:BS}
\end{table}

In search of candidates for spintronics applications,
two-dimensional (2D) topological insulators~\cite{KaneMele,KaneMele2,Ber06},
also known as quantum spin Hall systems, have emerged as a promising option.
In the topologically nontrivial phase,
such systems exhibit a gapped bulk, but gapless counterpropagating, ``helical'', edge states, \figr{edge}.
These counterpropagating edge states form Kramers pairs and are ensured by time-reversal symmetry (TRS).
They provide robust one-dimensional (1D) transport channels; an ideal edge has quantized conductance of $e^2/h$.
This quantization of conductance is the most direct experimental signature of the topological insulator phase
and has been successfully observed in HgTe/CdTe~\cite{Mole,Koenig,Roth09} and InAs/GaSb\cite{Kne11,Du15} quantum wells.

Central to spintronics applications is a question of {\em backscattering} (BS) effects,
which decrease the conductance from its ideal value $e^2/h$.
BS processes necessarily involve the change of the numbers $N_\pm$ of electrons
occupying the two branches of counterpropagating states, which we label $\pm$, \figr{edge}.
In symmetry terms, BS processes require a mechanism of breaking of $\Ux(1)$ symmetry with respect to rotations
about the $z$ axis in the effective {\em isospin}-$\f12$ space of the Kramers doublet $\pm$.

However, broken U(1) symmetry is only {\em necessary but not sufficient} to generate BS when TRS is preserved.
In particular, TR-symmetric disorder in a noninteracting system
generally breaks U(1) symmetry, yet does not result in BS:
the single-particle states completely evade localization even for stronger TR-symmetric disorder, and conductance remains ideal.
This feature constitutes the concept of topological protection of topological insulators by TRS.
The absence of BS in a noninteracting TR-symmetric system
is related to the conservation of the {\em single-particle energy},
and so, violation of this condition is another requirement for BS under preserved TRS.

One apparent effect that provides nonconservation of the single-particle energy is electron interactions.
One can therefore generally expect {\em inelastic} BS in the presence of
a TR-symmetric, U(1)-asymmetric single-particle backscattering field (commonly known as {\em Rashba spin-orbit coupling})
and TR-symmetric, U(1)-symmetric non-backscattering interactions
(other notable studied mechanisms of BS are magnetic moments~\cite{Mac09,Ta11}, phonons~\cite{Budich12},
and charge puddles~\cite{Vay13,Vay14,Vay16}).
Indeed, most theoretical works~\cite{Stroem,Sch12,Oreg,Crepin12,GeissCrep14,GeissCrep15,Kain14}
on this subject predicted that BS is generated by this combination, using either fermionic perturbation theory or renormalization group,
or a combination of bosonization technique and renormalization group.
On the other hand, a seemingly contradictory conclusion was reached in a recent work~\cite{F},
where it was argued that in the presence of the BS field and interactions
the system can be mapped onto an inhomogeneous Luttinger liquid, which implies that no BS is generated.

Motivated to resolve this potential controversy,
we perform an analysis of BS processes
due to such combination of TR-symmetric
U(1)-asymmetric, single-particle field (generalized Rashba spin-orbit coupling)
and TR-symmetric, U(1)-symmetric interactions,
considering quite general forms of them, to be defined precisely in \secr{model}.
We study the problem using fermionic perturbation theory and
calculate the amplitudes of BS processes and the associated BS currents.
We find that BS processes are crucially sensitive to the {\em locality} properties in real space
of the BS field and electron interactions,
to the {\em symmetry} of the latter in the $\pm$ isospin space,
as well as to the presence or absence of the {\em cutoff} of the electron spectrum.
In Ref.~\ocite{F},
the conclusion of absent BS was reached under quite stringent assumptions:
(i) local BS field, to be explained in \secr{model};
(ii.a) local (interaction potential is a delta function) {\em and} (ii.b) SU(2)-symmetric interactions;
(iii) absent cutoff.

Under these conditions, we do recover the result of Ref.~\ocite{F}, absent BS, within our perturbative approach.
In fact, on the one hand, we conjecture that
less restrictive conditions on the form of interactions are sufficient for the absence of BS:
they need to be just (ii.a) local {\em or} (ii.b) SU(2)-symmetric.
On the other hand, we find that violation of any
of these more general conditions leads to BS, manifested in finite BS amplitudes and currents.
Namely, if the BS field is nonlocal [(i) is violated],
BS is present even for local {\em or} SU(2)-symmetric interactions and absent cutoff.
If interactions are finite-range {\em and} SU(2)-asymmetric
[(ii.a) {\em and} (ii.b) are violated],
BS is present even for local BS field and absent cutoff.
If the cutoff is present [(iii) is violated],
BS is present even for local BS field and local {\em or} SU(2)-symmetric interactions.
The cases are summarized in \tabr{BS}.

We also explain that the latter cutoff effect reconciles the results
of Ref.~\ocite{F} with those of Refs.~\ocite{Stroem,Crepin12,GeissCrep14},
which predicted finite BS from an identical local fermionic model using the bosonization technique.
The bosonization procedure necessarily implies a finite cutoff of the electron spectrum,
and thus should produce BS even if the BS field and interactions are explicitly local in real space.

We illustrate the locality and symmetry conditions
for BS by deriving the scaling of the BS current with the bias voltage at zero temperature.

The rest of the paper is organized as follows.
In \secr{model}, we introduce the model.
In \secr{genexprs}, we derive the general expressions for BS amplitudes
and associated BS current to lowest orders in the BS field and interactions.
In \secr{genprops}, we establish the general locality, symmetry, and cutoff conditions
for the presence or absence of BS.
In \secr{scaling}, we derive the low-energy scaling behavior of the BS current.
Concluding remarks are presented in \secr{conclusion}.

\section{Model \lbl{sec:model}}

\begin{figure}
\includegraphics[width=.45\textwidth]{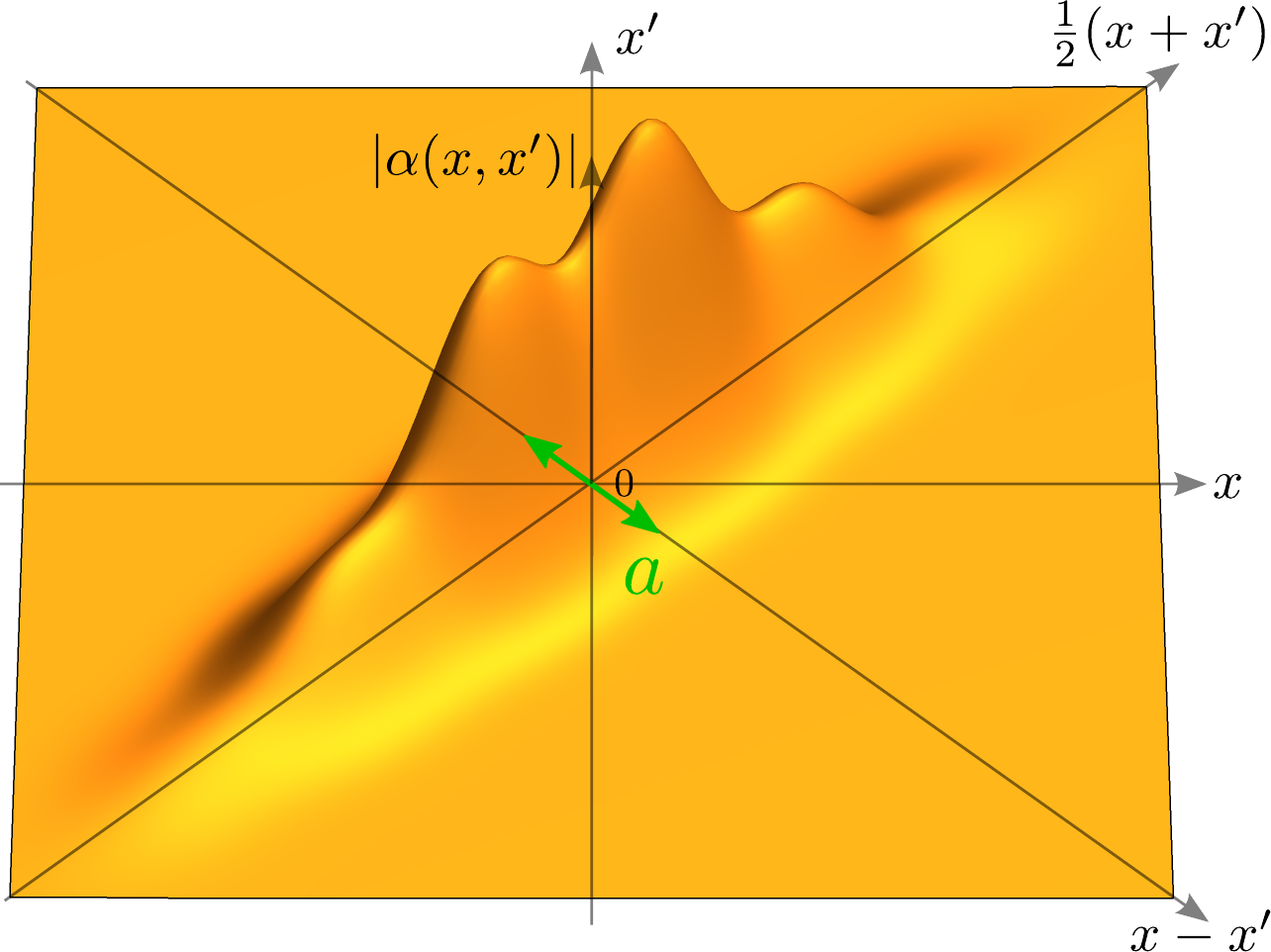}
\caption{
3D schematic of the nonlocal BS field $\al(x,x')$ [\eq{HR}].
}
\lbl{fig:al3D}
\end{figure}

We model the helical edge of a 2D topological insulator by the following many-body second-quantized Hamiltonian,
\begin{widetext}
\begin{align}
    \Hh&=\Hh_0+\de\Hh,
\lbl{eq:H}\\
    \de\Hh&=\Hh_\Rx+\Hh_\ix,
\lbl{eq:dH}\\
    \Hh_0 &= \sum_{\eta=\pm}\int \dx x \psih^\dg_\eta(x) (-\eta v \ix\partial_x)\psih_\eta(x)
        =\sum_{\eta=\pm} \int\f{\dx k}{2\pi} \varepsilon_{\eta}(k) \ch^\dg_{\eta k} \ch_{\eta k},
\lbl{eq:H0}\\
    \Hh_\Rx& =\int \dx x \dx x' \alpha(x,x')\left[ \ix\pd_x \psih^\dg_+(x) \psih_-(x') + \psih^\dg_+(x) (-\ix\pd_{x'}) \psih_-(x') \right]
        + \hc
\lbl{eq:HR}\\
    &= \int \f{\dx k \dx k'}{(2\pi)^2} \al_{k,k'}(k+k') \ch^\dg_{+,k}\ch_{-,k'} + \hc,\nn
\\
%
    \Hh_\ix&=\Hh_\ix^{(2)}+\Hh_\ix^{(4)},
\lbl{eq:Hi}\\
    \Hh_\ix^{(2)} &=\frac{1}{2} \int \dx x_1 \dx x_2\,
        g^{(2)}(x_1,x_2)
            [\psih^\dg_+(x_1) \psih^\dg_-(x_2) \psih_-(x_2)\psih_+(x_1)
            +\psih^\dg_-(x_1) \psih^\dg_+(x_2) \psih_+(x_2)\psih_-(x_1)]
\lbl{eq:Hi2}\\
    & = \f12\int\Dx k\,
        g^{(2)}_{k_1-k_1',k_2'-k_2} (\ch^\dg_{+,k_1} \ch^\dg_{-,k_2} \ch_{-,k_2'} \ch_{+,k_1'}
            +\ch^\dg_{-,k_1} \ch^\dg_{+,k_2} \ch_{+,k_2'} \ch_{-,k_1'}),  \nn\\
    \Hh_\ix^{(4)} &= \frac{1}{2} \int \dx x_1 \dx x_2\,
        g^{(4)}(x_1,x_2)
        [\psih^\dg_+(x_1) \psih^\dg_+(x_2) \psih_+ (x_2)\psih_+(x_1)+
        \psih^\dg_-(x_1) \psih^\dg_-(x_2) \psih_- (x_2)\psih_-(x_1)]
\lbl{eq:Hi4}\\
    &= \f12\int\Dx k\,
        g^{(4)}_{k_1-k_1',k_2'-k_2}
        (\ch^\dg_{+,k_1} \ch^\dg_{+,k_2} \ch_{+,k_2'} \ch_{+,k_1'}
        +\ch^\dg_{-,k_1} \ch^\dg_{-,k_2} \ch_{-,k_2'} \ch_{-,k_1'}).\nn
\end{align}
\end{widetext}
Here and below, we denote $\Dx k=\f{\dx k_1\dx k_2 \dx k_1'\dx k_2'}{(2\pi)^4}$ for brevity.

In \eq{H0}, $\Hh_0$ is the unperturbed Hamiltonian of an ideal non-interacting helical liquid,
characterized by the linear spectrum
\[
    \eps_\eta(k)=\eta vk
\]
with velocity $v$ of electrons of two chiralities $\eta=\pm$.
The fermion field operators
\beq
    \psih_\eta(x)=\int\f{\dx k}{2\pi} \ex^{\ix kx} \ch_{\eta k}
\lbl{eq:psi}
\eeq
are expanded in the plane-wave single-particle basis.
The fermion annihilation $\ch_{\eta k}$ and creation $\ch_{\eta k}^\dg$
operators in the plane-wave basis obey the anticommutation relations
\[
    \ch_{\eta k} \ch_{\eta' k'}+\ch_{\eta' k'}\ch_{\eta k}=0,
\]
\beq
    \ch_{\eta k} \ch^\dg_{\eta' k'}+\ch^\dg_{\eta' k'}\ch_{\eta k}= 2\pi\de(k-k')\delta_{\eta \eta'}.
\lbl{eq:cc}
\eeq
Accordingly, the field operators satisfy
\[
    \psih_\eta(x)\psih_{\eta'}(x')+\psih_{\eta'}(x')\psih_\eta(x)=0,
\]
\beq
    \psih_\eta(x)\psih_{\eta'}^\dg(x')+\psih_{\eta'}^\dg(x')\psih_\eta(x)=\de_{\eta\eta'}\de(x-x').
\lbl{eq:psipsi}
\eeq

{\em BS field.}
Next, $\Hh_\Rx$ [\eq{HR}] is the Hamiltonian of the single-particle {\em BS field} that couples $\pm$ states,
thereby breaking U(1) symmetry and changing the numbers
\beq
    \Nh_\eta=\int\f{\dx k}{2\pi} \ch^\dg_{\eta,k}\ch_{\eta,k}
\lbl{eq:Npm}
\eeq
of electrons in $\eta=\pm$ states.
We consider the most general form of such field allowed by TRS, characterized by a complex function
$\al(x,x')$ of two coordinates with the Fourier transform
\beq
    \al_{k,k'}=\int \dx x \dx x'\, \alpha(x,x') \ex^{\ix(-x k+x'k')}.
\lbl{eq:al}
\eeq
TRS imposes the constraint
\beq
    \al(x,x')=\al(x',x)
    \LRa \al_{k,k'}=\al_{-k',-k}.
\lbl{eq:TRS}
\eeq

The physical meaning of the dependence of the BS field $\al(x,x')$
on two coordinates is better elucidated in terms of the variables $x_+=\f12(x+x')$ and $x_-=x-x'$.
The dependence on $x_+=\f12(x+x')$ describes the spatial inhomogeneity of the BS field;
the case of $\al(x,x')=\al(x-x')$ independent of $x_+$
is the translationally invariant case.

The dependence on $x_-$ describes the {\em nonlocality} of the BS field;
generally, we assume $\al(x,x')$ to be a decaying function of $x_-$
over the microscopic spatial scale $a$ of the low-energy theory (see below).
The form $\al(x,x')=\al_0(x_+)\de(x_-)$, where $\de(x_-)$ is a delta function,
is the familiar conventional {\em local} form of the BS field,
known as Rashba spin-orbit coupling, studied earlier~\cite{Stroem,Sch12,Oreg,Crepin12,GeissCrep14,GeissCrep15,Kain14}.
When the dependence on $x_-$ has a finite extent,
the field is nonlocal since in this case the first-quantized operator
is an {\em integral} operator, and the dependence on $x_-$ describes the {\em kernel} of this operator.
Nonlocal form of the BS field can be regarded as the {\em generalization} of the Rashba spin-orbit coupling.

Nonlocality of $\al(x,x')$, i.e., deviation from the form $\al_0(x_+)\de(x_-)$
can be rephrased in terms of the derivatives of the electron field of higher order than linear:
since the low-energy electron fields $\psih(x)$
vary over spatial scales much larger than the extent $a$ of $\al(x_+,x_-)$ in $x_-$,
one may perform in $\Hh_\Rx$ the Taylor expansion of the fields $\psih(x)=\psih(x_++\f12x_-)$ and $\psih(x')=\psih(x_+-\f12x_-)$
in $x_-$ about the common center $x_+$, to arrive at an equivalent form of $\Hh_\Rx$,
which would be {\em local} (a single integral over $x_+$), but contain derivatives of higher order than linear.
The integral form of $\Hh_\Rx$ is, however, more practically convenient for our general analysis.

{\em Interactions.}
Next, $\Hh_\ix$ [\eqss{Hi}{Hi2}{Hi4}] are the two-particle interactions.
To clearly distinguish between the two ingredients necessary for BS, mentioned in \secr{intro},
breaking of U(1) symmetry and nonconservation of the single-particle energy,
we consider $\Ux(1)$-symmetric interactions.
Such interactions preserve $N_\pm$ numbers and by themselves do not result in BS:
the fermion model $\Hh_0+\Hh_\ix$ maps~\cite{Wu06,Xu06}
onto the Luttinger liquid model of free bosons
for arbitrary strength of interactions,
and exhibits~\cite{MaslovStone,SafiSchulz,Ponomarenko} ideal conductance $e^2/h$.
And so, in our model, the single-particle BS field $\Hh_\Rx$ is responsible for breaking U(1) symmetry,
while U(1)-symmetric interactions $\Hh_\ix$ are responsible for nonconservation of the single-particle energy.

We consider the U(1)-symmetric interactions of the density-density type,
for which the electron densities $\psih^\dg_\eta(x)\psih_\eta(x)$ of $\eta=\pm$ states
interact with themselves ($n=4$) and each other
($n=2$, we label interactions according to ``$g$-ology'' convention~\cite{Solyom79,GiaSchu,Gia})
via the finite-range potentials $g^{(n)}(x_1,x_2)$, with Fourier transforms
\begin{align*}
    g^{(n)}_{q_1,q_2}&=\int \dx x_1 \dx x_2\, g^{(n)}(x_1,x_2) \ex^{\ix(-x_1q_1+x_2q_2)}.
\end{align*}
Naturally, the interaction potentials are symmetric, $g^{(n)}(x_1,x_2)=g^{(n)}(x_2,x_1)$;
equivalently, in momentum space, $g^{(n)}_{q_1,q_2}=g^{(n)}_{-q_2,-q_1}$.
These interactions become SU(2)-symmetric when $g^{(2)}(x_1,x_2)=g^{(4)}(x_1,x_2)$.
We mention that the form of U(1)-symmetric interactions we consider is still not the most general one;
however, our conclusions are not sensitive to this assumption.

Similarly to the properties of $\al(x,x')$,
the dependence of $g^{(n)}(x_1,x_2)$ on $x_+=\f12(x_1+x_2)$ describes the spatial inhomogeneity of interactions;
translationally invariant interactions $g^{(n)}(x_1,x_2)=g^{(n)}(x_1-x_2)$ depend only on the difference $x_-=x_1-x_2$.
We consider spatially inhomogeneous interactions
to demonstrate that our conclusions do not rely on the assumption of translational symmetry of interactions.
The dependence on $x_-$ describes the interaction potential,
assumed to be a decaying function.
For {\em local}, also called {\em contact}, interactions, the potential is a delta function,
$g^{(n)}(x_1,x_2)=g^{(n)}(x_+)\de(x_-)$.
Local interactions between fermions have a special property
that, due to the Pauli exclusion principle, their structure in the isospin space
is effectively restricted. In particular, for {\em two-component} fermions,
as is the case for the helical states $\pm$ we consider here,
there is effectively {\em only one} coupling constant.
Indeed, in the most general isospin structure $\sum_{\eta_1\eta_2\eta_1'\eta_2'}g^{\eta_1\eta_1'}_{\eta_2\eta_2'}(x)
\psih^\dg_{\eta_1}(x)\psih^\dg_{\eta_2}(x)\psih_{\eta_2'}(x)\psih_{\eta_1'}(x)$
of local interactions,
all terms containing particle-particle bilinears $\psih_\eta(x)\psih_\eta(x)=0$ with the same label $\eta$
and their conjugates vanish due to fermion anticommutation relations \eqn{cc}.
The remaining nonvanishing terms can all be brought to the form
$\psih^\dg_+(x)\psih^\dg_-(x)\psih_-(x)\psih_+(x)$
of $g^{(2)}$-type interactions [\eq{Hi2}].
And so, this is the most general {\em effective} form of local interactions between two-component fermions,
characterized by {\em just one} coupling constant.

{\em Cutoff.}
We also consider the effect of the cutoff of the electron spectrum,
which we formalize as follows. Consider a {\em cutoff function} $F_k$, \figr{F},
which is a nongrowing function of momentum $|k|$
with a maximum $F_0=1$ at the Fermi level, $k=0$ in our case,
and vanishing at $k\rarr\pm\iy$. The range of $F_k$, the cutoff scale,
is determined by the inverse $1/a$ of the microscopic spatial scale $a$ of the low-energy theory;
in our case, $a$ is set by the lateral extent of the microscopic wave functions of the helical edge states
and related to the gap $\sim v/a$ of the bulk electron spectrum.

The cutoff is introduced by considering the modified Hamiltonian $\Hth=\Hh[\cth]$,
obtained by replacing the electron operators $\ch_{\eta,k}$
in the original Hamiltonian $\Hh=\Hh[\ch]$ [\eq{H}] by the ``constrained'' operators
\beq
    \cth_{\eta,k}=\sq{F_k}\ch_{\eta,k},
\lbl{eq:ct}
\eeq
\beq
    \psith_\eta(x)
    =\int\f{\dx k}{2\pi} \ex^{\ix kx} \cth_{\eta k}
    =\int\f{\dx k}{2\pi} \ex^{\ix kx} \sq{F_k} \ch_{\eta k}.
\lbl{eq:psit}
\eeq
In such Hamiltonian, momentum integrations
are constrained to the range of the bandwidth $\sim1/a$ due to the presence of the cutoff function,
whereas for absent cutoff ($F_k\equiv 1$) the integrations are unrestricted.

For our purpose, an important observation is that constraining the electron spectrum
in momentum space by introducing a cutoff
leads to effective nonlocality of the electron operators in real space.
This is most easily illustrated by the anticommutation relations
\beq
    \cth_{\eta k} \cth^\dg_{\eta' k'}+\cth^\dg_{\eta' k'}\cth_{\eta k}= 2\pi\de(k-k')\delta_{\eta \eta'} F_k,
\lbl{eq:cc}
\eeq
\beq
    \psith_\eta(x)\psith_{\eta'}^\dg(x')+\psith_{\eta'}^\dg(x')\psith_\eta(x)=\de_{\eta\eta'}F(x-x'),
\lbl{eq:psitpsit}
\eeq
\[
    F(x)=\int\f{\dx k}{2\pi}\ex^{\ix k x} F_k,
\]
for the modified electron operators. Instead of the delta function $\de(x-x')$
(due to the completeness of the plane-wave system $\{\ex^{\ix k x}\}_k$ with all continuous $k$)
for absent cutoff on the right hand side of the anticommutation relations \eqn{psipsi},
the inverse Fourier transform $F(x-x')$ of the cutoff function $F_k$ is present, which naturally
has a finite range determined by the microscopic spatial scale $a$.
This nonlocality manifests in other properties as well.
Thus, one can anticipate the effect of the cutoff to be qualitatively similar to the effect of explicitly
nonlocal BS field or interactions, as we prove to indeed be the case.

\begin{figure}
\includegraphics[width=.45\textwidth]{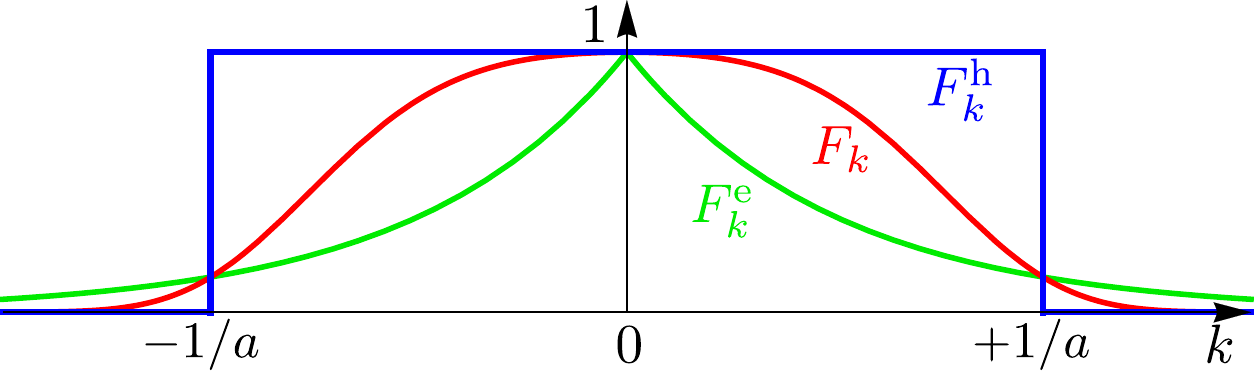}
\caption{
Examples of cutoff functions: (a) generic smooth $F_k$ (red), (b) hard $F_k^\hx$ [\eq{Fh}, blue], and (c) exponential $F_k^\ex$ [\eq{Fe}, green].
Different smoothness properties of the cutoff function lead to different scaling behaviors of the BS current, see \secsr{yescutoff}{scalingconclusion}.
}
\lbl{fig:F}
\end{figure}

We consider a general ``smooth'' cutoff, when $F_k$ is continuously differentiable,
as well as two types of commonly used cutoffs, \figr{F}.
One is the ``hard'' cutoff
\beq
    F_k^\hx=\lt\{\ba{c} 1,\spc |k|<1/a, \\ 0,\spc |k|>1/a,\ea\rt.
    \spc
    F^\hx(x)=\f{\sin \f{x}{a}}{\pi x},
\lbl{eq:Fh}
\eeq
described by a rectangular window function with a bandwidth $1/a$.
In this case, the electron states outside of the band $[-1/a,+1/a]$
are completely discarded and the electron Hilbert space is restricted.

The other cutoff we use is exponential,
\beq
    F_k^\ex=\ex^{-|k|a},
    \spc
    F^\ex(x)=\f1\pi\f{a}{x^2+a^2}.
\lbl{eq:Fe}
\eeq
The fermion exponential cutoff is closely related to the exponential cutoff
commonly used in bosonization~\cite{Gia,Delft,Gogolin}.
In particular, the anticommutation relation for the electron field operators
(as well as the electron Green function)
obtained from bosonization formalism has exactly the form of \eq{psitpsit} with $F^\ex(x)$ [\eq{Fe}] on the right hand side.

All these three generalizations
(nonlocal BS field, finite-range SU(2)-asymmetric interactions, and finite cutoff)
we consider here should be present in a realistic low-energy model.
A cutoff is present (even if implicitly) in a continuous low-energy theory
since its Hilbert space is, by construction, restricted to the states of interest.
Further, at a general level, the structure of a low-energy model is guided solely by the
true symmetries of the system; in our case, TRS is the only such symmetry.
Locality of operators is not required by any symmetry or any other fundamental reason.
As a result, any nonlocal operator (of single-particle or interaction type) that satisfies
just TRS is admitted in the low-energy model.
The ``bare'' operators may be local (as in the case of the BS field, Rashba coupling)
or have higher symmetry than the true one (the matrix elements of the bare SU(2)-spin-symmetric Coulomb interactions
are SU(2)-symmetric in the space of the Kramers doublet $\pm$).
However, additional nonlocal terms and terms that lower the symmetry down to the true one get generated
via virtual transitions to the high-energy electron states that
are not included in the low-energy single-particle Hilbert space of the edge states
(the bulk states, in the case of the topological insulator).
Such processes set the spatial nonlocality scale, which is given by the microscopic scale $a$ of the theory (see above);
technically, it is determined by the extent of the correlation functions of the high-energy states.
Note that, through this mechanism, nonlocality of a low-energy theory is also essentially a consequence of restricting the Hilbert space.
And though additional nonlocal and symmetry-restoring terms
may be of smaller magnitude than the bare ones,
they are nonetheless admitted and can affect physics in a qualitative way, as we demonstrate here.

For establishing the general properties of BS in \secsr{genexprs}{genprops},
no further assumptions about the forms of the BS field, interactions,
and cutoff are necessary.
For calculating the scaling behavior of the BS current in \secr{scaling},
a few more specific assumptions will be made.

\section{Backscattering processes and current \lbl{sec:genexprs}}

\begin{figure}
\includegraphics[width=.48\textwidth]{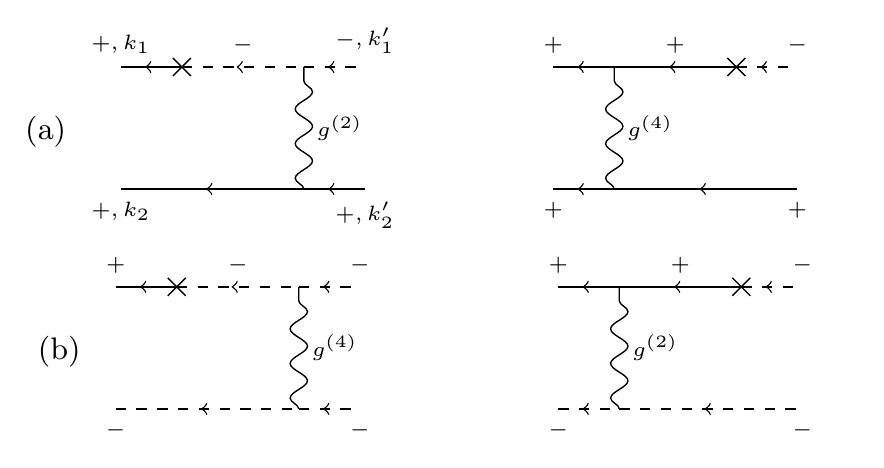}
\caption{
Diagrams for one-particle backscattering (1PBS) processes,
to the lowest, first, order in the BS field and interactions.
The BS field is represented by crosses and electron interactions by wiggly lines.
The diagrams (a) and (b) describe the BS amplitudes $\Tc[^{+-}_{++}|^{k_1k_1'}_{k_2k_2'}]$ [\eqs{Tc1a}{T1a}]
and $\Tc[^{+-}_{--}|^{k_1k_1'}_{k_2k_2'}]$ [\eqs{Tc1b}{T1b}], respectively.
}
\lbl{fig:diag1}
\end{figure}

\begin{figure}
\includegraphics[width=.45\textwidth]{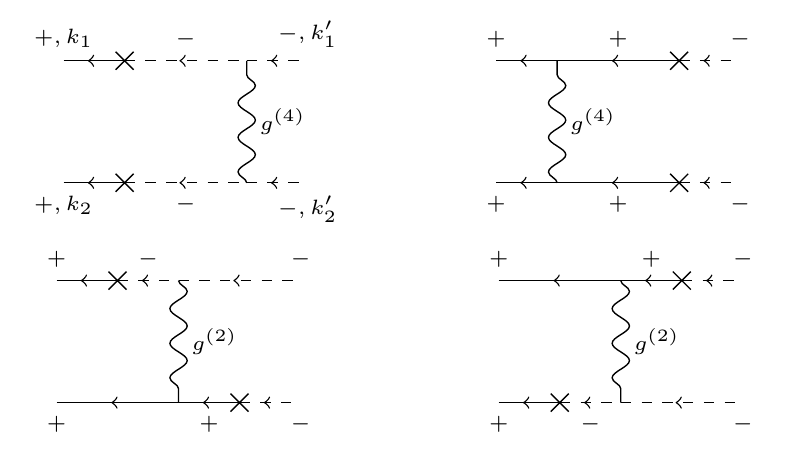}
\caption{
Diagrams for two-particle backscattering (2PBS) processes,
with the amplitude $\Tc[^{+-}_{+-}|^{k_1k_1'}_{k_2k_2'}]$ [\eqs{Tc2}{T2}],
to the second order in the BS field and first order in interactions.
}
\lbl{fig:diag2}
\end{figure}

We study BS processes, which involve changes of $N_\pm$ [\eq{Npm}] numbers of electrons in the two edge-state branches $\pm$
due to the cooperative effect of the BS field $\Hh_\Rx$ and electron interactions $\Hh_\ix$.
To specify the problem, we calculate the {\em rate of change} of the numbers $N_\pm$,
commonly referred to as the {\em ``BS current''}.
Within the framework of the Fermi ``golden rule''~\cite{LLIV}, the BS current reads
\beq
    \de I =\sum_{f} 2\pi \de(E_f-E_i) |\lan f|\Tc|i\ran|^2 \de N_{fi}.
\lbl{eq:dIdef}
\eeq
Here, $|i\ran=|i,N_{+i},N_{-i}\ran$ and $|f\ran=|f,N_{+f},N_{-f}\ran$
are the initial and final states, which are eigenstates of the unperturbed Hamiltonian $\Hh_0$ [\eq{H0}],
with energies $E_i$ and $E_f$ and conserved numbers $N_{\pm i}$ and $N_{\pm f}$, respectively
(the total number $N_{+i}+N_{-i}=N_{+f}+N_{-f}$ of electrons is always conserved for the full Hamiltonian $\Hh$);
$\de N_{fi}=N_{+f}-N_{+i}=N_{-i}-N_{-f}$ is the backscattered charge in a BS process.

Next, $\lan f|\Tc|i\ran$ in \eq{dIdef} is the scattering amplitude between the initial $|i\ran$ and final $|f\ran$ states.
It is defined through the scattering matrix
\beq
    \Sc=\Tx\exp\lt[-\ix\int_{-\iy}^{+\iy}\dx t\, \de\Hth(t)\rt]
    ,\spc
\lbl{eq:S}
\eeq
\[
    \de\Hth(t)=\ex^{\ix\Hh_0 t}\de\Hth\ex^{-\ix\Hh_0 t}
    ,\spc
    \de\Hth=\de\Hh[\cth],
\]
as
\beq
    \lan f|\Sc|i\ran=\lan f|i\ran -\ix 2\pi\de(E_f-E_i)\lan f|\Tc|i\ran.
\lbl{eq:T}
\eeq

We assume that both the BS field $\Hh_\Rx$ and interactions $\Hh_\ix$
are weak and can therefore be treated perturbatively.
We consider one- and two-particle backscattering (1PBS and 2PBS) processes $\lan N_++\de N,N_--\de N|\Tc|N_+,N_-\ran$
with the change of electron numbers $\de N=1,2$.
The calculations of the scattering amplitudes are done using the standard diagrammatic technique~\cite{LLIV,AGD}.

We first remind that without interactions no BS processes are possible:
due to energy conservation in \eq{dIdef}, single-particle BS
could occur only between the single-particle states $|+,+k\ran$ and $|-,-k\ran$ of the same energy $vk$;
such states form a Kramers pair and the matrix element
\[
    \lan-,-k|\Tc|+,+k\ran=F_k \al_{-k,k}(-k+k)=0
\]
of the TRS-preserving perturbation $\Hh_\Rx$ between them vanishes.
This explicitly demonstrates that TRS precludes any elastic BS, in which the single-particle energy is conserved.
(Note that this result relies on having {\em only one} Kramers pair of states at a given energy
and so is valid only for one pair of counterpropagating states~\cite{Koenig,Wu06,Xu06}.)

The conservation of the single-particle energy holds no more in the presence of interactions and BS processes become possible,
since the nonvanishing matrix elements $\sq{F_k F_{k'}}\al_{k,k'}(k+k')\neq0$
with unrelated $k$ and $k'$ can now enter the scattering amplitude $\Tc$.

1PBS processes with $\de N=1$ first arise in the first order in the BS field and first order in interactions;
the scattering amplitudes for the elementary processes between two-particle states
are represented by the diagrams in \figr{diag1} and read
\begin{widetext}
\begin{align}
    \Tc[^{+-}_{++}|^{k_1k_1'}_{k_2k_2'}]&=T[^{+-}_{++}|^{k_1k_1'}_{k_2k_2'}]-T[^{+-}_{++}|^{k_2k_1'}_{k_1k_2'}],
\lbl{eq:Tc1a}\\
     T[^{+-}_{++}|^{k_1k_1'}_{k_2k_2'}]
    &=\Fc^{k_1k_1'}_{k_2k_2'}
    \f1v  \int\f{\dx q}{2\pi} \lt(g^{(2)}_{q,k_2'-k_2} \al_{k_1,k'_1+q} F_{k_1'+q} -g^{(4)}_{q,k_2'-k_2} \al_{k_1-q,k'_1} F_{k_1-q}\rt),
\lbl{eq:T1a}
\end{align}
\begin{align}
    \Tc[^{+-}_{--}|^{k_1k_1'}_{k_2k_2'}]&=T[^{+-}_{--}|^{k_1k_1'}_{k_2k_2'}]-T[^{+-}_{--}|^{k_1k_2'}_{k_2k_1'}],
\lbl{eq:Tc1b}\\
    T[^{+-}_{--}|^{k_1k_1'}_{k_2k_2'}]
    &=\Fc^{k_1k_1'}_{k_2k_2'}
    \f1{v} \int\f{\dx q}{2\pi} \left(
        g^{(4)}_{q,k_2'-k_2} \al_{k_1,k'_1+q} F_{k_1'+q} -g^{(2)}_{q,k_2'-k_2} \al_{k_1-q,k'_1} F_{k_1-q} \right),
\lbl{eq:T1b}
\end{align}
\beq
    \Fc^{k_1k_1'}_{k_2k_2'}=\sq{F_{k_1}F_{k_1'}F_{k_2}F_{k_2'}}.
\lbl{eq:Fc}
\eeq
Here and below, the labels of the initial and final two-particle states
are placed according to the 2D layout of the diagrams in \figsr{diag1}{diag2}.
These processes represent BS of a single electron
accompanied by a particle-hole excitation on $+$ or $-$ branch;
this mechanism has been identified in Refs.~\ocite{Sch12,Kain14,Oreg}.

As a technical point, the terms in the BS field linear in momentum [\eq{HR}] cancel
exactly the denominator of the single-particle causal Green function
\beq
    \Gt_\pm(\e,k)=\f{F_k}{\e-\eps_\pm(k)+\ix\,\sgn(\eps_\pm(k)\pm V/2)}.
\lbl{eq:Gt}
\eeq
Note that the Green function involves $F_k$ in the presence of the cutoff
and see \eq{f} below for the occupation of electron states.

The 2PBS processes with $\de N=2$ first arise in the second order in the BS field and first order in interactions;
the scattering amplitudes for the elementary processes between two-particle states
are represented by the diagrams in \figr{diag2} and read
\beq
    \Tc[^{+-}_{+-}|^{k_1k_1'}_{k_2k_2'}]
    =\f12(
    T[^{+-}_{+-}|^{k_1k_1'}_{k_2k_2'}]
    -T[^{+-}_{+-}|^{k_2k_1'}_{k_1k_2'}]
    -T[^{+-}_{+-}|^{k_1k_2'}_{k_2k_1'}]
    +T[^{+-}_{+-}|^{k_2k_2'}_{k_1k_1'}]),
\lbl{eq:Tc2}
\eeq
\[
    T[^{+-}_{+-}|^{k_1k_1'}_{k_2k_2'}]
    =\Fc^{k_1k_1'}_{k_2k_2'}
    \frac{1}{v^2}
        \int\f{\dx q_1\dx q_2}{(2\pi)^2} \left[
        g^{(4)}_{q_1,q_2} (\al_{k_1,k_1'+q_1} \al_{k_2,k_2'-q_2} F_{k_1'+q_1} F_{k_2'-q_2}
        +\al_{k_1-q_1,k_1'} \al_{k_2+q_2,k_2'} F_{k_1-q_1}F_{k_2+q_2})
        \rt.
\]
\beq
    \lt.
        -2g^{(2)}_{q_1,q_2}\al_{k_1,k_1'+q_1} \al_{k_2+q_2,k_2'} F_{k_1'+q_1} F_{k_2+q_2}
    \right].
\lbl{eq:T2}
\eeq
\end{widetext}
These processes describe simultaneous transfer of two electrons between $-$ and $+$ branches;
this mechanism has been identified in Ref.~\onlinecite{Crepin12}.
In \eqss{T1a}{T1b}{T2}, the combination \eqn{Fc} of the cutoff functions arises
from external momenta $k_{1,2},k_{1,2}'$,
while individual cutoff functions $F_{\ldots}$ come from the Green functions \eqn{Gt} in the diagrams.

We point out that the scattering amplitudes
are antisymmetric with respect to the transposition of the same-isospin electrons in the final or initial two-particle states.
This is a consequence of the Fermi statistics of electrons.
As we demonstrate below,
this antisymmetry property in conjunction with the locality properties of interactions
has profound consequences for BS.

We assume that the initial unperturbed eigenstate $|i\ran$ of $\Hh_0$ is the current-carrying state
described by the zero-temperature Fermi distribution function
\beq
    f_{\pm,k}=\lt\{\ba{c} 0, \spc \varepsilon_{\pm}(k) \pm V/2>0,\\
        1, \spc \varepsilon_{\pm}(k) \pm V/2<0.\ea\rt.
\lbl{eq:f}
\eeq
where the bias voltage $V$ {\em in energy units}
represents the shift of the chemical potentials of $+$ and $-$ states.
The resulting contributions $\de I_{1,2}$ to the BS current [\eq{dIdef}]
\[
    \de I =\de I_1+\de I_2
\]
due to 1PBS processes \eqsn{Tc1a}{Tc1b} and 2PBS processes \eqn{Tc2}, respectively, read
\begin{widetext}
\beq
    \delta I_1=\delta I_1[^{+-}_{++}]+\delta I_1[^{+-}_{--}]-\delta I_1[^{-+}_{++}]-\delta I_1[^{-+}_{--}],
\lbl{eq:dI1}
\eeq
\begin{align}
\label{eq:H1pIbs}
    \delta I_1[^{+-}_{++}] &
    =\int \Dx k\,
        \f{2\pi}v\delta(+k_1+k_2+k'_1-k'_2)
        \lt|\Tc[^{+-}_{++}|^{k_1k_1'}_{k_2k_2'}]\rt|^2
        (1-f_{+,k_1})(1-f_{+,k_2})f_{+,k'_2}f_{-,k'_1},
        \notag \\
    \delta I_1[^{+-}_{--}]& =
    \int\Dx k\,
    \f{2\pi}v\delta(+k_1-k_2+k'_1+k'_2)
        \left|\Tc[^{+-}_{--}|^{k_1k_1'}_{k_2k_2'}]\right|^2
        (1-f_{+,k_1})(1-f_{-,k_2})f_{-,k'_2} f_{-,k'_1}, \notag \\
    \delta I_1[^{-+}_{++}] & =
    \int\Dx k\,
    \f{2\pi}v\delta(-k_1+k_2-k'_1-k'_2)
        \left|\Tc[^{-+}_{++}|^{k_1k_1'}_{k_2k_2'}]\right|^2
        (1-f_{-,k_1})(1-f_{+,k_2})f_{+,k'_2} f_{+,k'_1}, \notag \\
    \delta I_1[^{-+}_{--}] & =
    \int\Dx k\,
    \f{2\pi}v\delta(-k_1-k_2-k'_1+k'_2)
        \left|\Tc[^{-+}_{--}|^{k_1k_1'}_{k_2k_2'}]\right|^2
        (1-f_{-,k_1})(1-f_{-,k_2})f_{-,k'_2}f_{+,k'_1}, \notag
\end{align}
\beq
    \Tc[^{-+}_{++}|^{k_1k_1'}_{k_2k_2'}]=\Tc[^{+-}_{++}|^{k_1'k_1}_{k_2'k_2}]^*
,\spc
    \Tc[^{-+}_{--}|^{k_1k_1'}_{k_2k_2'}]=\Tc[^{+-}_{--}|^{k_1'k_1}_{k_2'k_2}]^*,
\lbl{eq:Tc1r}
\eeq
\beq
    \de I_2=\delta I_2[^{+-}_{+-}]-\delta I_2[^{-+}_{-+}],
\lbl{eq:dI2}
\eeq
\begin{align}
    \delta I_2[^{+-}_{+-}]
        &=2
        \int\Dx k\,
        \f{2\pi}v\delta(+k_1+k_2+k_1'+k_2') (1-f_{+,k_1})(1-f_{+,k_2})f_{-,k_2'} f_{-,k_1'}
        \left|\Tc[^{+-}_{+-}|^{k_1k_1'}_{k_2k_2'}]\right|^2 ,
\notag \\
    \delta I_2[^{-+}_{-+}]
    &=2
    \int\Dx k\,
    \f{2\pi}v\delta(-k_1-k_2-k_1'-k_2') (1-f_{-,k_1})(1-f_{-,k_2}) f_{+,k_2'} f_{+,k_1'}
    \left| \Tc[^{-+}_{-+}|^{k_1k_1'}_{k_2k_2'}]\right|^2,
\notag
\end{align}
\beq
    \Tc[^{-+}_{-+}|^{k_1k_1'}_{k_2k_2'}]=\Tc[^{+-}_{+-}|^{k_1'k_1}_{k_2'k_2}]^*.
\lbl{eq:Tc2r}
\eeq
\end{widetext}
Note that the amplitudes of the ``reversed'' processes [\eqs{Tc1r}{Tc2r}] are related to each other by hermitian conjugation.
This is the consequence of the optical theorem stemming from the unitarity of the scattering matrix $\Sc$ [\eq{S}]
and the fact that the amplitudes are calculated to the lowest order in perturbations.

For $V>0$, the integration regions in \eqs{dI1}{dI2}
are such that $\de I_1[^{-+}_{++}]=\de I_1[^{-+}_{--}]=\de I_2[^{-+}_{-+}]=0$
vanish and the contributions come from
$\de I_1[^{+-}_{++}]=\de I_1[^{+-}_{--}]$ and $\de I_2[^{+-}_{+-}]$ only.
Using the properties of the delta functions due to the energy conservation and the step functions \eqn{f}
of the Fermi distributions, the integrals can be further brought to the forms
\begin{align}
    \de I_1[^{+-}_{++}]
    &=\f1v
    \f1{2^2(2\pi)^3}
        \int_{-k_V}^0\dx k_+
        \int_{-k_V-k_+}^{+k_V+k_+}\dx k_-\nn\\
        &\tm\int_{-k_V+k_+}^{-k_V-k_+}\dx k_+'\,
        |\Tc[^{+-}_{++}|^{k_1k_1'}_{k_2k_2'}]|_{k_++k_-'=0}|^2,
\lbl{eq:dI1*}
\end{align}
\begin{align}
    \de I_2[^{+-}_{+-}]
    &=\f2v
    \f1{2^2(2\pi)^3}
        \int_{-k_V}^{+k_V}\dx k_+\,
        \int_{-k_V-k_+}^{+k_V+k_+}\dx k_-\nn\\
        &\tm\int_{-k_V+k_+}^{+k_V-k_+}\dx k_-'\,
        |\Tc[^{+-}_{+-}|^{k_1k_1'}_{k_2k_2'}]|_{k_++k_+'=0}|^2.
\lbl{eq:dI2*}
\end{align}
Here
\[
    k_V=\f{V}{v}
\]
is the momentum scale set by the bias voltage,
\beq
    k_\pm=k_1\pm k_2, \spc k_\pm'=k_1'\pm k_2'
\lbl{eq:kconv}
\eeq
are the convenient variables, and the amplitudes are taken momenta satisfying the energy conservations,
$k_++k_-'=0$ and $k_++k_+'=0$, respectively.

\section{General locality, symmetry, and cutoff properties\lbl{sec:genprops}}

The key consequences for BS stemming from the locality of the BS field,
locality and symmetry of interactions,
and the cutoff
follow directly from the BS amplitudes.

Consider a local BS field, $\al_{k,k'}=\al_{k-k'}$ and absent cutoff, $F_k\equiv1$,
such that all momentum integrations are unconstrained.
The BS amplitudes \eqsn{T1a}{T2} [and similar for \eq{T1b}] then take the form
\begin{align*}
    T[^{+-}_{++}|^{k_1k_1'}_{k_2k_2'}]
    =\f1{v} \int\f{\dx q}{2\pi} \lt( g^{(2)}_{q,k_2-k'_2} - g_{q,k_2-k'_2}^{(4)} \rt) \al_{k_1-k'_1-q},\\
\end{align*}
\begin{align*}
    T[^{+-}_{+-}|^{k_1k_1'}_{k_2k_2'}]
    =&\f2{v^2}
        \int\f{\dx q_1\dx q_2}{(2\pi)^2}\\
        &\tm
        \lt(g^{(4)}_{q_1,q_2}-g^{(2)}_{q_1,q_2}\rt) \al_{k_1-k_1'-q_1} \al_{k_2-k_2'+q_2}.
\end{align*}
We see that interactions enter as the difference $g^{(4)}_{q_1,q_2}-g^{(2)}_{q_1,q_2}$,
and so, the 1PBS and 2PBS amplitudes vanish (even before antisymmetrization)
for any form (finite-range or local) of SU(2)-symmetric interactions, $g^{(4)}_{q_1,q_2}=g^{(2)}_{q_1,q_2}$.

Now, consider both BS field $\al_{k,k'}=\al_{k-k'}$ and interactions $g^{(4),(2)}_{q_1,q_2}=g^{(4),(2)}_{q_1-q_2}$ local,
but present cutoff. As a direct consequence of the locality of the BS field and interactions,
the integrands, except for the cutoff functions,
of the contributions to the full antisymmetrized amplitudes \eqsn{Tc1a}{Tc2}
can be made identical by changing the integration momenta.
For $g^{(4)}$ interactions, also the cutoff functions $F$... become the same
and the respective contribution to the full antisymmetrized amplitudes vanish exactly;
this is a manifestation of the fact, explained in \secr{model}, that local interactions
between fermions of the same component $\pm$ are effectively absent, also in the presence of the cutoff.
For $g^{(2)}$ interactions, however, the cutoff-function parts are different,
which results in the net forms
\begin{widetext}
\beq
    \Tc[^{+-}_{++}|^{k_1k_1'}_{k_2k_2'}]
    =\Fc^{k_1k_1'}_{k_2k_2'} \f1v \int\f{\dx q}{2\pi}(F_{k_1'+q}-F_{k_1'+q+k_2-k_1})
    g^{(2)}_{q+k_2-k'_2} \al_{k_1-k'_1-q},
\lbl{eq:Tc1al}
\eeq
\beq
    \Tc[^{+-}_{+-}|^{k_1 k_1'}_{k_2 k_2'}]
    =\Fc^{k_1k_1'}_{k_2k_2'}\f{-1}{v^2}
        \int\f{\dx q_1 \dx q_2}{(2\pi)^2}
            (F_{k_1'+q_1}-F_{k_1'+q_1+k_2-k_1})
            (F_{k_2+q_2}-F_{k_2+q_2+k_1'-k_2'})
        g^{(2)}_{q_1-q_2}
        \al_{k_1-k_1'-q_1} \al_{k_2-k_2'+q_2}.
\lbl{eq:Tc2l}
\eeq
\end{widetext}
So, for local BS field and interactions
the anti-symmetrized BS amplitudes vanish for absent cutoff, $F_k\equiv 1$;
however, they generally do not vanish in the presence of the cutoff.

Combining the above results,
let us summarize the conditions under which we have obtained vanishing of the BS amplitudes:
(i) local form of the BS field;
(ii.a) local {\em or} (ii.b) SU(2)-symmetric interactions;
(iii) absent cutoff of the electron spectrum.

Comparing these results with those of Ref.~\ocite{F},
we note that there the conclusion of absent BS was reached for (ii.a) local {\em and} (ii.b) SU(2)-symmetric interactions,
whereas we obtain that BS amplitudes vanish already when {\em only one} of the two conditions is satisfied.
We thus {\em conjecture} that BS is absent for a more general form of interactions, namely,
when they are {\em either} (ii.a) local (in which case the symmetry question is irrelevant, since
their structure is effectively unique, as explained in \secr{model}) {\em or}
(ii.b) SU(2)-symmetric (in which case they could be finite-range).
We point out the caveat that this is indeed a conjecture,
since our perturbative analysis proves vanishing of the BS amplitudes to a given order only.
To prove the absence of BS in general, a nonperturbative method is required.

However, when {\em any one} of  these three conditions (with relaxed conditions on the form of interactions)
is violated, BS amplitudes are nonzero
and BS is present; perturbative analysis {\em is sufficient} to prove this point.
Explicitly, if the BS field is nonlocal [(i) is violated],
BS is present even for local {\em or} SU(2)-symmetric interactions [(ii.a) {\em or} (ii.b) is satisfied]
and absent cutoff [(iii) is satisfied].
If interactions are finite-range {\em and} SU(2)-asymmetric
[(ii.a) {\em and} (ii.b) are violated],
BS is present even for local BS field [(i) is satisfied] and absent cutoff [(iii) is satisfied].
If the cutoff is present [(iii) is violated],
BS is present even for local BS field [(i) is satisfied] and local {\em or} SU(2)-symmetric interactions [(ii.a) {\em or} (ii.b) is satisfied].
The cases are summarized in \tabr{BS}.

Our results thus establish precise conditions for the absence of BS,
and thus, for the validity of the conclusions of Ref.~\ocite{F}.
We see that these conditions are quite stringent.
As explained in detail in \secr{model},
all of these effects are generally present in the proper low-energy model for the helical edge states of a 2D topological insulator
and will provide BS of some magnitude.
We quantify the magnitude of BS due to these effects in the next section by calculating the scaling dependence of the BS current.

\section{Low-energy scaling\lbl{sec:scaling}}

\begin{figure}
\includegraphics[width=.45\textwidth]{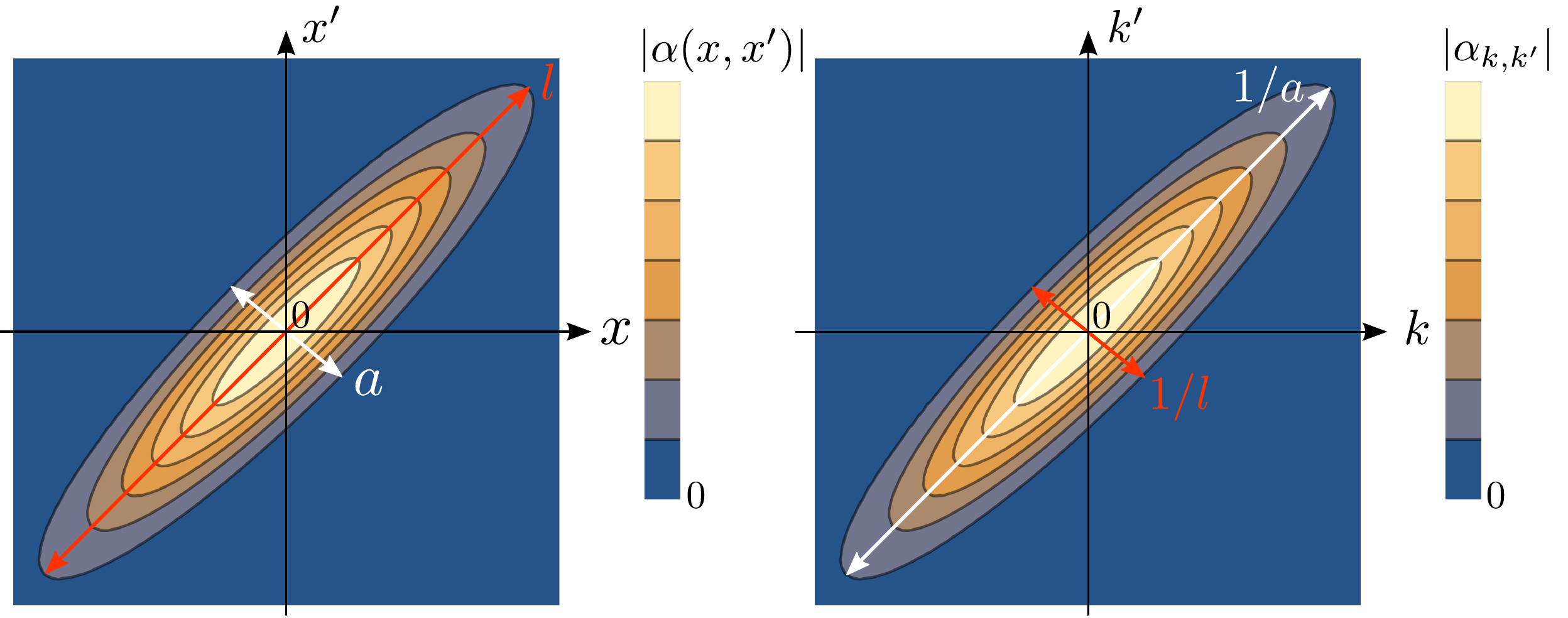}
\caption{Contour plots of the BS field $\al(x,x')$ and its Fourier transform $\al_{k,k'}$.
}
\lbl{fig:al2D}
\end{figure}

\begin{figure}
\includegraphics[width=.45\textwidth]{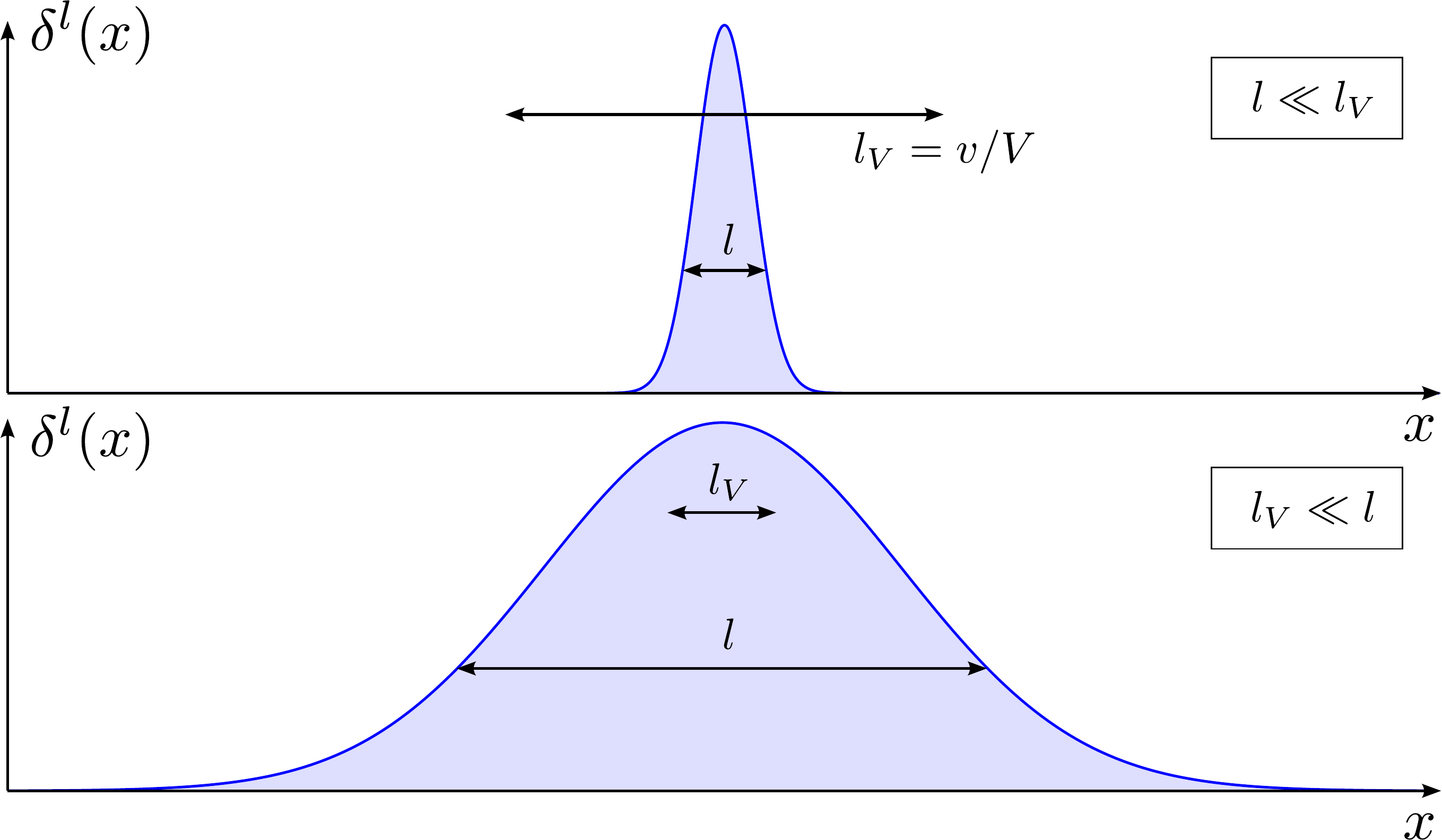}
\caption{
The profile $\de^l(x)$ of the BS field \eqn{alfactorx}
and a real-space illustration of the two regimes $V\ll V_l$ and $V_l\ll V$
for the bias voltage. The bias voltage length scale $l_V=1/k_V=v/V$
is much larger ($l\ll l_V$) and smaller ($l_V\ll l$)
than the extent $l$ of the BS field profile, respectively.
}
\label{fig:Vregimes}
\end{figure}

We now calculate the asymptotic behavior of the BS currents $\de I_{1,2}(V)$ [\eqs{dI1}{dI2}]
due to 1PBS and 2PBS processes in the limit of small bias $V$.
To make the results more physically transparent,
we make the following additional natural assumptions
about the forms of the BS field and interactions.

\subsection{Forms of the BS field and interactions}

As explained in \secr{model},
the dependence on the variable $\f12(x+x')$ determines
the spatial ``profile'' of the BS field $\al(x,x')$,
whereas the dependence on $x-x'$ determines the shape of the nonlocal kernel.
To clearly distinguish these two different aspects,
for the calculations below, we consider a factorized form
\beq
    \al(x,x')=\de^l(\tf12(x+x'))\altd(x-x').
\lbl{eq:alfactorx}
\eeq
Here, $\de^l(\f12(x+x'))$ is a dimensionless function describing the profile,
which we assume to have a spatial extent $l$, such that $\de^l(0)=1$ at its maximum, \figr{Vregimes};
$\altd(x-x')$ is the integral kernel of the microscopic extent $a$.
Since $a$ is the microscopic scale of the theory, naturally we assume $l\gtrsim a$.
Accordingly, in momentum space,
\begin{align}
    \al_{k,k'}= 2\pi\de^l_{k-k'}\altd_{\f12(k+k')}.
\lbl{eq:alfactor}
\end{align}
Here, the Fourier transform $\de^l_q=\int\dx x\,\ex^{-\ix q x} \de^l(x)$
is a peaked function in momentum space centered around $q=0$ with height $\de^l_{q=0}\sim l$ and extent $1/l$.
Is it a quasi delta function, $\int\f{\dx q}{2\pi} 2\pi \de^l_q=\de^l(x=0)=1$,
which becomes the true delta function if the profile $\de^l(x)\equiv 1$
is translationally invariant in real space.

The Fourier transform $\altd_{(k+k')/2}$ of the nonlocal kernel $\altd(x-x')$ has a large extent $1/a$ in momentum space.
We also assume that the kernel $\altd(x-x')$ decays sufficiently fast,
so that $\altd_q$ is analytic at $q=0$ in momentum space.
For the form \eqn{alfactor} of the BS field, the TRS constraint \eqn{TRS}
means that $\altd(x-x')=\altd(x'-x)$ and $\altd_q=\altd_{-q}$ are even.
And so, the expansion at small momenta $qa\ll1$ contains only even powers and reads
\beq
    \altd_q=\altd_0+\tf12\pd^2\altd_0 q^2+\Oc(q^4).
\lbl{eq:altdexp}
\eeq
This small-$q$ expansion allows us to further elucidate the locality properties of the BS field.
If the BS field kernel $\altd(x-x')=\altd_0\de(x-x')$ is local in real space,
its Fourier transform is exactly constant $\altd_q\equiv\altd_0$.
So, for a nonlocal BS field kernel, $\altd_0$ may be referred to as the ``local part'',
while higher-order terms in $q$ describe the ``nonlocal part'' of the BS field kernel.
Note that this representation also illustrates the point made in \secr{model}
that the nonlocal integral form can be reformulated in the local form with higher-order
derivatives of the electron fields, since momentum powers correspond
to derivatives in real space, $q\rarr-\ix\pd_x$.

Below, we also consider translationally invariant electron interactions, $g^{(n)}(x_1,x_2)=\gt^{(n)}(x_1-x_2)$,
for which $g^{(n)}_{q,q'}=2\pi\de(q-q') \gt^{(n)}_q $. The expansions
\beq
    \gt^{(n)}_q
    =\gt^{(n)}_0+\tf12\pd^2 \gt^{(n)}_0 q^2 + \Oc(q^4)
\lbl{eq:gexp}
\eeq
of the Fourier transforms of the interaction potentials at small momenta $qa\ll 1$
[where we assume that the range of the interaction potentials $\gt^{(n)}(x_1-x_2)$
is also equal to the microscopic scale $a$]
contains only even powers due to the interchange symmetry of interactions.
Similarly to the BS field \eqn{altdexp} above, $\gt^{(n)}_0$ describe the local parts of the interactions,
which would be the only part for truly local interactions, $\gt^{(n)}(x_1-x_2)=\gt^{(n)}_0\de(x_1-x_2)$,
while the higher-order terms in $q$ describe their nonlocal parts.

We calculate the BS current in the limit of the bias voltage
$V\ll v/a$ much smaller than the microscopic energy scale, set by the bulk gap.
At the same time, the relation between the bias $V$ and the energy scale
\beq
    V_l=\f{v}{l}
\lbl{eq:Vl}
\eeq
associated with the spatial extent $l\gtrsim a$ of the BS field profile may be arbitrary.
For short extent $l\sim a$ of the BS field, a microscopic-scale ``impurity'',
there is only one regime $V\ll V_l$.
However, for large enough spatial extent $l\gg a$ of the BS field,
an additional regime $V_l\ll V \ll v/a$ exists, \figr{Vregimes}.
We derive the general expressions for arbitrary $V/V_l$,
and then calculate the asymptotic scaling behaviors in the limits $V\ll V_l$ and $V_l\ll V$ (when the latter exists).
As bias $V$ is swiped, one crosses over between the two regimes at $V\sim V_l$.

As follows from \eqs{dI1*}{dI2*}, the Fermi functions \eqn{f} together with
the delta function due to the energy conservation
confine the external momenta $k_{1,2},k_{1,2}'\lesssim k_V=V/v$ of the scattering amplitudes
to the region of size set by the bias voltage.

\subsection{Nonlocal BS field or interactions, absent cutoff \lbl{sec:nocutoff}}

Here, we consider nonlocal BS field or interactions and absent cutoff, $F_k\equiv1$.
Given the properties of the BS field and interactions assumed above,
the kernel $\altd_q$ and interaction potentials $\gt^{(n)}_q$, which have a large momentum extent $1/a$,
may be expanded in external momenta in the BS amplitudes for $V\ll v/a$,
whereas the BS field profile $\de^l_{\ldots}$ stays as is for arbitrary $V/V_l$.
The expanded expressions for the 1PBS~[\eq{Tc1a}, and similar for \eq{Tc1b}] and 2PBS~[\eq{Tc2}] amplitudes read
\begin{widetext}
\beq
    \Tc[^{+-}_{++}|^{k_1k_1'}_{k_2k_2'}]
    =\f1{v} 2\pi\de^l_{k_+-k_+'}
        \lt[ \tf12\pd^2(\gt^{(2)}_0-\gt^{(4)}_0)(k_+'-k_-'-k_+)\altd_0
            +\gt^{(2)}_0 \tf12 \pd^2\altd_0 k_+'  \rt] k_-,
\lbl{eq:T1exp}
\eeq
\beq
    \Tc[^{+-}_{+-}|^{k_1k_1'}_{k_2k_2'}]
    =\frac{1}{v^2} k_- k_-'
    \De^l_{k_+-k_+'},
\lbl{eq:T2exp}
\eeq
\beq
    \De^l_\ka
    =\int\f{\dx q}{2\pi}
    2\pi \de^l_{\tf{\ka}2-q} 2\pi\de^l_{\tf{\ka}2+q}
        \lt[
        -\pd^2(\gt^{(4)}_q-\gt^{(2)}_q)
            \altd_{\f12q}^2
        -\pd(\gt^{(4)}_q-\gt^{(2)}_q) 2 \pd \altd_{\f12q} \altd_{\f12q}
        +\gt^{(2)}_q\pd\altd_{\f12q}^2
    \rt],
\lbl{eq:De}
\eeq
in terms of the convenient variables \eqn{kconv}.
In deriving these expressions, we used the symmetry relation $\altd_{q}=\altd_{-q}$ due to TRS,
and the relations $\pd\altd_q=-\pd\altd_{-q}$ and $\pd\altd_0=0$ that follow.

The resulting general expressions for the BS currents [\eqs{dI1*}{dI2*}],
providing the leading asymptotics for arbitrary $V/V_l$ for bias $V\ll v/a$, read
\beq
    \de I_1(V)
    =\de I_1[^{+-}_{++}]+\de I_1[^{+-}_{--}]
    =\f2v
    |A|^2
    \f1{2^2(2\pi)^3}
        \int_{-k_V}^0\dx k_+
        \int_{-k_V-k_+}^{+k_V+k_+}\dx k_-\, k_-^2
        \int_{-k_V+k_+}^{-k_V-k_+}\dx k_+'\,  k_+'^2
        (\f1v 2\pi)^2|\de^l_{k_+-k_+'}|^2,
\lbl{eq:dI1exp}
\eeq
\[
    A=\tf12\pd^2(\gt^{(2)}_0-\gt^{(4)}_0)\altd_0+\gt^{(2)}_0 \tf12 \pd^2\altd_0,
\]
\beq
    \de I_2(V)
    =\de I_2[^{+-}_{+-}]
    =\f2v\f1{2^2(2\pi)^3}
        \int_{-k_V}^{+k_V}\dx k_+\,
        \f1{v^4}|\De^l_{2k_+}|^2
        \int_{-k_V-k_+}^{+k_V+k_+}\dx k_- \,k_-^2
        \int_{-k_V+k_+}^{+k_V-k_+}\dx k_-'\, k_-'^2.
\lbl{eq:dI2exp}
\eeq
\end{widetext}

We see that locality properties of the BS field and interactions
and the isospin symmetry of the latter explicitly manifest,
in accord with the general conclusions of \secr{genprops}:
in this form, nonlocality enters as derivatives
$\pd(\gt^{(2)}_q-\gt^{(4)}_q)$ and $\pd\altd_q$ of the Fourier transforms,
as explained around \eqs{altdexp}{gexp}.

The coefficient $A$ and function $\De^l_\ka$, and consequently, the BS currents $\de I_{1,2}$, are zero only if
(i) BS field is local ($\pd\altd_q=0$) and (ii.a) the interactions are local ($\pd \gt^{(n)}_q=0$)
{\em or} (ii.b) SU(2)-symmetric ($\gt^{(2)}_q=\gt^{(4)}_q$).
Otherwise, $A$, $\De^l_\ka$, and the currents $\de I_{1,2}$ are nonzero.

In the regimes $V\ll V_l$ and $V_l\ll V$ (when the latter exists), further approximations are possible
and the scaling dependence on the bias voltage $V$ can be obtained explicitly.
In real space, the two regimes correspond to the bias length scale $l_V=1/k_V=v/V$
being much larger ($l\ll l_V$) and smaller ($l_V\ll l$) than the extent $l$
of the BS field profile, respectively, \figr{Vregimes}.
The regime $V_l\ll V$ can be regarded as ``quasi translationally invariant'',
with the peaked ``quasi delta function'' $\de^l_\ka$ representing near momentum conservation.

The peaked functions $\de^l_\ka$ and (consequently) $\De^l_\ka$ depend on the difference $\ka=k_+-k_+'$ between
the total final $k_+$ and initial $k_+'$ momenta of the BS process and have extent $1/l$.
In the regime $V\ll V_l$, the integration range $\sim k_V$ for all involved momenta is much smaller than this extent
and, to the leading order, these functions may be taken at zero momentum.
In the opposite regime $V_l\ll V$, $\de^l_\ka$ and $\De^l_\ka$
are restricted to a small region in $\ka$ compared to the integration range $\sim k_V$
and are sharply peaked; as a result, the integration region is effectively constrained.
For the scaling dependencies of the BS currents with the bias voltage, we obtain, to the leading order,
\begin{align}
    &\de I_1(V)
    =\f2v|A|^2\f1{(2\pi)^3}
    \nn\\
    &\tm\lt\{\ba{l}
    \f{11}{630} k_V^7(\f1v2\pi)^2|\de^l_0|^2
    \propto V^7, \spc V\ll V_l,\\
    \f{11}{11520} k_V^6
        \int_{-\iy}^{+\iy} \dx k_+'\, (\f1v 2\pi)^2|\de^l_{-k_+'}|^2\propto V^6, \spc V_l\ll V,
    \ea\rt.
\lbl{eq:dI1scal}
\end{align}
\begin{align}
    &\de I_2(V)
    =\f2{v^5}\f1{(2\pi)^3} \nn\\
    &\tm\lt\{\ba{l}
    \f{32}{315} k_V^7|\De^l_0|^2
    \propto V^7, \spc V\ll V_l,\\
        \f19 k_V^6
        \int_{-\iy}^{+\iy}\dx k_+\,
        |\De^l_{2k_+}|^2 \propto V^6, \spc V_l\ll V.
    \ea\rt.
\lbl{eq:dI2scal}
\end{align}

\begin{figure}
\includegraphics[width=.4\textwidth]{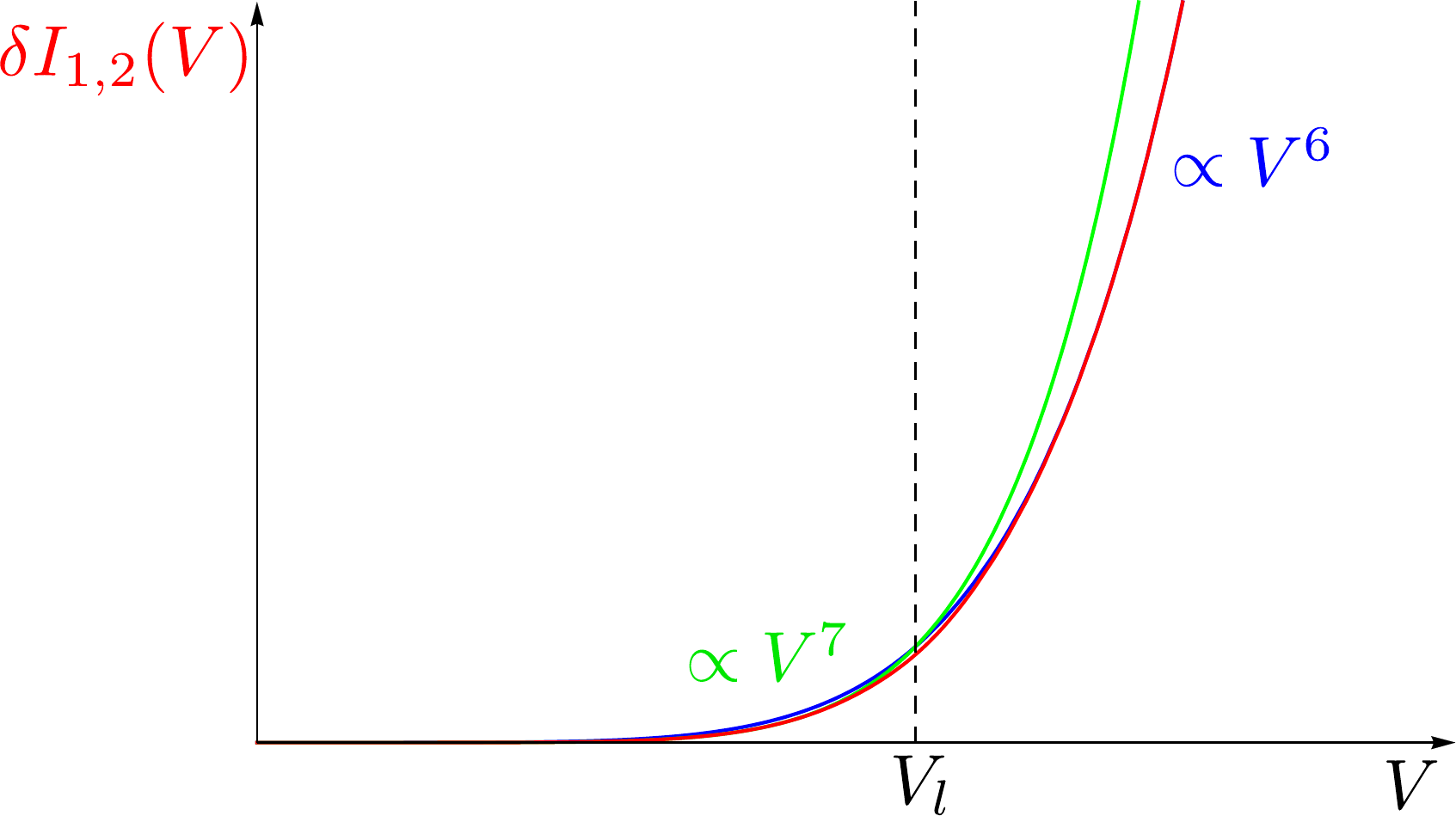}
\caption{
Schematic dependence of the backscattering currents $\de I_{1,2}(V)$ (red)
for nonlocal BS field or nonlocal and SU(2)-asymmetric interactions [\eqs{dI1scal}{dI2scal}],
or in the presence of a smooth cutoff of the electron spectrum [\eqs{dI1Fsm}{dI2Fsm}].
There are two scaling regimes,
$\de I_{1,2}(V)\propto V^7$ (green) at $V\ll V_l$
and
$\de I_{1,2}(V)\propto V^6$ (blue) at $V_l\ll V$,
and a crossover between them at $V\sim V_l$,
where $V_l=v/l$ [\eq{Vl}] is the energy scale set by the spatial extent $l$
of the BS field profile.
}
\label{fig:crossover}
\end{figure}

We see that the scaling is the same for 1PBS and 2PBS
currents in each regime, but differs between two regimes:
$\de I_{1,2}(V)\propto V^7$ for  $V\ll V_l$ and $\de I_{1,2}(V)\propto V^6$ for $V_l\ll V$.
At $V\sim V_l$, both expressions become parametrically the same and a crossover between the two regimes occurs;
the dependence is shown schematically in Fig.~\ref{fig:crossover}.
We discuss the scaling properties further in \secr{yescutoff}.

The dependence of $\de I_{1,2}(V)$ on the extent $l$ of the BS field is determined
by the functions $\de^l_\ka$ and $\De^l_\ka$. Since
\[
    \de^l_0\propto l,\spc \int_{-\iy}^{+\iy}\dx\ka\,|\de^l_\ka|^2\propto l,
\]
for the 1PBS current, we obtain
\[
    \de I_1(V)\propto\lt\{ \ba{c} l^2,\spc V\ll V_l,\\
        l,\spc V_l\ll V.\ea \rt.
\]
Note that, in the regime $V_l\ll V$, the 1PBS current is {\em extensive},
proportional to the size $l$ of the region where the BS field exists.

For $l\gg a$, the dependence of the 2PBS current $\de I_2(V)$ on $l$ varies with the locality and symmetry of interactions.
For $l\gg a$, the integral in $\De^l_\ka$ [\eq{De}] is constrained to the region $\sim1/l$ around $q=0$,
much smaller than the extent $1/a$ of $\altd_q$ and $\gt_q^{(n)}$,
which allows us to expand them in the integrand.
For finite-range SU(2)-asymmetric interactions, to the leading order,
\[
    \De^l_\ka
    \approx-\int\f{\dx q}{2\pi}
    2\pi \de^l_{\tf{\ka}2-q} 2\pi\de^l_{\tf{\ka}2+q}
        \pd^2(\gt^{(4)}_0-\gt^{(2)}_0)
            \altd_0^2,
\]
\[
    \De^l_0\propto l, \spc \int_{-\iy}^{+\iy}\dx\ka\,|\De^l_\ka|^2
    \propto l,
\]
and the 2PBS current
\[
    \de I_2(V)\propto\lt\{ \ba{c} l^2,\spc V\ll V_l,\\
        l,\spc V_l\ll V,\ea \rt.
\]
has the same dependence on $l$ as the 1PBS current $\de I_1(V)$.
For local or SU(2)-symmetric interactions, however,
the above expression vanishes. In the next order,
\[
    \De^l_\ka
    \approx \int\f{\dx q}{2\pi}
    2\pi \de^l_{\tf{\ka}2-q} 2\pi\de^l_{\tf{\ka}2+q}
        \gt^{(2)}_0 (\pd^2 \altd_0 \tf{q}2)^2,
\]
\[
    \De^l_0\propto \f1l,\spc
    \int_{-\iy}^{+\iy}\dx\ka\,|\De^l_\ka|^2
    \propto \f1{l^3},
\]
and the 2PBS current
\[
    \de I_2(V)\propto\lt\{ \ba{c} 1/l^2,\spc V\ll V_l,\\
        1/l^3,\spc V_l\ll V,\ea \rt.
\]
decays with the extent $l$ of the BS field in both regimes.

\subsection{Local BS field and interactions, present cutoff \lbl{sec:yescutoff}}

Here we consider local BS field and interactions, but present cutoff.
For the local BS field, $\al_q$ describes the Fourier transform of the spatial profile.
As in \secr{nocutoff}, we assume that the profile has extent $l$ in real space;
in terms of \eqs{alfactorx}{alfactor},
$
    \al_q=2\pi \de^l_q \altd_0,
$
with a constant $\altd_0$ and profile $\de^l_q$.
Assuming that interactions are also translationally invariant, $g_q^{(2)}=2\pi\de(q)\gt_0^{(2)}$,
the BS amplitudes \eqsn{Tc1al}{Tc2l} reduce to
\begin{widetext}
\beq
    \Tc[^{+-}_{++}|^{k_1k_1'}_{k_2k_2'}]
    =\Fc^{k_1k_1'}_{k_2k_2'} \f1v (F_{k_1'+k_2'-k_2}-F_{k_1'+k_2'-k_1})
    \gt^{(2)}_0 \al_{k_1+k_2-k'_1-k_2'},
\lbl{eq:T1F}
\eeq
\beq
    \Tc[^{+-}_{+-}|^{k_1 k_1'}_{k_2 k_2'}]
    =\Fc^{k_1k_1'}_{k_2k_2'}\f1{v^2}
        \int\f{\dx q}{2\pi}
            (F_{\f12(k_-+k_+')+q}-F_{\f12(-k_-+k_+')+q})
            (F_{\f12(k_++k_-')+q}-F_{\f12(k_+-k_-')+q})
    \gt^{(2)}_0 \al_{\f12(k_+-k_+')-q} \al_{\f12(k_+-k_+')+q}.
\lbl{eq:T2F}
\eeq
\end{widetext}

We consider three types of the cutoff function $F_k$: smooth, hard [\eq{Fh}], and exponential [\eq{Fe}].

For a smooth cutoff function, with a continuous first derivative $\pd F_k$
at all momenta and a well-defined second derivative $\pd^2 F_0$ at $k=0$ ($\pd F_0=0$ since $k=0$ is a maximum),
the BS amplitudes, expanded to the lowest order in external momenta $k_-$ and $k_-'$, read
\beq
    \Tc[^{+-}_{++}|^{k_1k_1'}_{k_2k_2'}]
    =
    \f1v \f12\pd^2 F_0 k_-(2 k_+'-k_+)
    \gt^{(2)}_0 \al_{k_+-k_+'},
\lbl{eq:T1Fsm}
\eeq
\begin{align}
    \Tc[^{+-}_{+-}|^{k_1 k_1'}_{k_2 k_2'}]
    &=\f1{v^2}
    k_-k_-'\Det^l_{k_+-k_+'},\lbl{eq:T2Fsm}\\
    \Det^l_\ka&=
        \int\f{\dx q}{2\pi}
    (\pd F_q)^2
    \gt^{(2)}_0
    \al_{\f\ka2-q} \al_{\f\ka2+q}.\nn
\end{align}

The 2PBS amplitude has the same dependence on external momenta
as in the case  of nonlocal BS field or interactions and absent cutoff in \secr{nocutoff};
the function $\Det^l_\ka $ has the same main properties as $\De^l_\ka$ [\eq{De}]: extent $1/l$ and height $\propto l$.
The 1PBS amplitude has a slightly different dependence, but is still of the same, second order in external momenta.
As a result, the scaling dependence of the 1PBS and 2PBS current for smooth cutoff is the same as
in the cases of absent cutoff, but nonlocal BS field or interactions, studied in \secr{nocutoff},
\begin{align}
    \de I_1(V)
    &=\f2v\lt(\f1v\f12\pd^2 F_0 \gt^{(2)}_0 \rt)^2
    \f1{(2\pi)^3} \nn\\
    &\tm \lt\{\ba{l} \f1{20} k_V^7 |\al_{0}|^2 \propto l^2 V^7, \spc  V\ll V_l,\\
    \f{11}{11520} k_V^6 \int_{-\iy}^{+\iy} \dx k_+'\, |\al_{k_+-k_+'}|^2 \propto l V^6, \spc  V_l\ll V,\ea\rt.
\lbl{eq:dI1Fsm}
\end{align}
\begin{align}
    \de I_2(V)
    &=\f2{v^5} \f1{(2\pi)^3}
    \nn\\
    &\tm \lt\{\ba{l} \f{32}{315} k_V^7 |\Det^l_0|^2
    \propto l^2 V^7, \spc V\ll V_l, \\
        \f19 k_V^6
        \int_{-\iy}^{+\iy}\dx k_+\,
        |\Det^l_{2k_+}|^2 \propto l V^6, \spc V_l\ll V.
    \ea\rt.
\lbl{eq:dI2Fsm}
\end{align}

For the hard cutoff \eqn{Fh},
the BS amplitudes to the leading order in external momenta $k_{1,2},k_{1,2}'\ll 1/a$ read
\[
    \Tc[^{+-}_{++}|^{k_1k_1'}_{k_2k_2'}]\equiv 0,
\]
\begin{align}
    &\Tc[^{+-}_{+-}|^{k_1 k_1'}_{k_2 k_2'}]
    =\f1{v^2}\f1{2\pi}
        \gt^{(2)}_0 \al_{-1/a} \al_{+1/a} \nn\\
      &\tm  (|k_1-k_2'|+|k_2-k_1'|-|k_1-k_1'|-|k_2-k_2'|).
\lbl{eq:T2h}
\end{align}
The 1PBS amplitude vanishes since the cutoff function is piecewise constant.
For the 2PBS amplitude, the contributions come from small regions around the cutoff momenta $\pm1/a$.
The respective BS currents read
\beq
    \de I_1=0,
\lbl{eq:dI1h}
\eeq
\beq
    \de I_2
    =\lt(\f1{v^2}\f1{2\pi}\gt^{(2)}_0 \al_{-1/a} \al_{+1/a}\rt)^2
    \f1{(2\pi)^3}\f{810029}{2278125} k_V^5 \propto V^5.
\lbl{eq:dI2h}
\eeq
Since the BS field is taken at large momenta $\pm1/a$ and thus does not depend on small external momenta,
the dependence of $\de I_2(V)$ on $V$ is the same for any relation between $V$ and $V_l$.

For the exponential cutoff \eqn{Fe}, the 1PBS amplitude \eqn{T1F}
to the leading order in external momenta reads
\begin{align*}
    \Tc[^{+-}_{++}|^{k_1k_1'}_{k_2k_2'}]
    =&\f1v a(|k_1'+k_2'-k_1|-|k_1'+k_2'-k_2|)
    \\
    &\tm\gt^{(2)}_0 \al_{k_1+k_2-k'_1-k_2'}.
\end{align*}
The corresponding 1PBS current in the two regimes is
\begin{align*}
    \de I_1(V)
    =&\f2v
    \lt(\f1v a \gt^{(2)}_0\rt)^2
    \f1{(2\pi)^3} \\
    &\tm \lt\{\ba{l} \f{191}{11520} k_V^5 |\al_0^l|^2 \propto l^2 k_V^5, \\
        \f1{384} k_V^4 \int_{-\iy}^{+\iy} \dx k_+'\, |\al^l_{k_+-k_+'}|^2 \propto l k_V^4.
    \ea \rt.
\end{align*}

The 2PBS amplitude \eqn{T2F} for the exponential cutoff \eqn{Fe}
has the following asymptotic expressions in the two regimes,
to the leading order in $k_-$ and $k_-'$,
\begin{align*}
    \Tc[^{+-}_{+-}|^{k_1 k_1'}_{k_2 k_2'}]
    &=\f1{v^2}
    \tilde{\Det}^l_0 k_- k_-', \spc V\ll V_l, \nn\\
    \tilde{\Det}^l_0
    &=a^2
        \int_{-\iy}^{+\iy}\f{\dx q}{2\pi}
            \ex^{-2a|q|}
        \gt^{(2)}_0 \al_{-q} \al_{+q},
\end{align*}
\begin{align}
    \Tc[^{+-}_{+-}|^{k_1 k_1'}_{k_2 k_2'}]|_{k_++k_+'=0}
    =\f1{v^2}
        \sgn k_- \,\sgn k_-'
        d^l_{k_+-k_+'},     \spc V_l\ll V,
\lbl{eq:T2e} \\
        d^l_{k_+-k_+'}=
        4a^2\int_{-\iy}^{+\iy}\f{\dx q}{2\pi}\,
            q^2
    \gt^{(2)}_0 \al_{\f12(k_+-k_+')-q} \al_{\f12(k_+-k_+')+q},
\nn
\end{align}
where the latter is taken at momenta satisfying the energy conservation $k_++k_+'=0$.
The corresponding 2PBS current in the two regimes reads
\begin{align}
    \de I_2(V)
    &=\f2{v^5}
    \f1{(2\pi)^3}\nn\\
    &\tm\lt\{\ba{l}
        \f{32}{315} k_V^7 |\tilde{\Det}^l_0|^2 \propto l^2 V^7, \spc V\ll V_l,\\
        k_V^2 \int_{-\iy}^{+\iy}\dx k_+\,
        |d^l_{2k_+}|^2 \propto V^2/l^3,\spc V_l\ll V. \ea\rt.
\lbl{eq:dI2e}
\end{align}

Below we discuss the results of \secsr{nocutoff}{yescutoff}.

\subsection{Smoothness and scaling, bosonization\lbl{sec:scalingconclusion}}

Aggregating the results of \secsr{nocutoff}{yescutoff},
we observe that the low-bias scaling of the BS current is crucially sensitive to
the smoothness properties of the scattering amplitudes at small external momenta.
The latter, in turn, are determined by the smoothness properties of the BS field kernel, interactions,
and cutoff function at all momenta (only 1PBS amplitude is determined by those at small momenta).
When all three are smooth, the scaling of the BS currents $\de I_{1,2}(V)$ is quite universal:
it is the same whether the current is due to nonlocality of BS field or interactions [\eqs{dI1scal}{dI2scal}]
or to the presence of the cutoff [\eqs{dI1Fsm}{dI2Fsm}], see also \figr{crossover}.
This universality is due to the analytic dependence
of the antisymmetrized BS amplitudes [\eqs{T1exp}{T2exp} and \eqs{T1Fsm}{T2Fsm}]
at small momenta $k_-$ and $k_-'$.

However, for both hard [\eq{Fh}] and exponential [\eq{Fe}] cutoffs, the scaling is different.
Both of these cutoffs, though commonly used, are not everywhere differentiable:
the hard cutoff \eqn{Fh} has a jump at the band edges $k=\pm1/a$,
while the exponential cutoff \eqn{Fe} has a jump in the first derivative at $k=0$
(related to which is the power-law tail of the Lorenzian in real space).
As a consequence of these singularities,
the scattering amplitudes [\eqs{T2h}{T2e}] are not smooth at small momenta,
which results in the BS current scaling as a lower power of the bias voltage [\eqs{dI2h}{dI2e}].
Thus the low-bias scaling of the BS current depends on the smoothness properties of the cutoff function,
suggesting that the choice of cutoff is a nontrivial matter.
On the other hand, the BS field kernel and interaction potential
are generally naturally smooth at all momenta~\cite{Coulomb}.

In Refs.~\ocite{Stroem,Crepin12,GeissCrep14}, BS was studied using bosonization technique:
under the assumptions of local BS field and interactions
and absent cutoff in the {\em initial fermionic model},
finite BS was found in the bosonized version of this model.
This result was later questioned in Ref.~\ocite{F},
where no BS was found under the same assumptions about the fermionic model
(the structure of local interactions in the $\pm$ isospin space is effectively unique, see \secr{model}).
The latter result was obtained by finding the exact solution (subsequently recovered in Ref.~\ocite{Dolcini16})
to the single-particle problem for arbitrary local BS field.
In \secr{genprops}, we have confirmed perturbatively that BS is indeed absent in the fermionic model under these assumptions.
The bosonization-based results of Refs.~\ocite{Stroem,Crepin12,GeissCrep14}
are reconciled with these findings of the fermionic method as follows.
Bosonization procedure itself necessarily implies imposing a cutoff on the electron spectrum.
Consequently, even if the initial fermionic model has no cutoff, its bosonized version effectively contains one.
Hence, the bosonization-based results have to be compared
to those for the fermionic model {\em with a cutoff}
and there is no qualitative disagreement, since in the latter case BS is present.

\section{Summary\lbl{sec:conclusion}}

In this work, we have studied backscattering effects
in the presence of a single-particle time-reversal-symmetric backscattering field (generalized Rashba coupling)
and time-reversal-symmetric non-backscattering electron interactions in a 1D helical liquid.
We have found that backscattering of some magnitude is inevitable when either
the backscattering field or interactions are nonlocal, or when the electron spectrum has a finite cutoff;
precise conditions are summarized at the end of \secr{genprops} and in \tabr{BS}.
All of these effects that lead to backscattering are expected to be present in a real system.
We have quantified backscattering by calculating the scaling behavior of the backscattering current at low bias voltage.
We have found that scaling is sensitive to the smoothness properties of the backscattering field, interactions, and cutoff
in the momentum space, as discussed in \secr{scalingconclusion}.
We have also resolved the potential controversy between Refs.~\ocite{F} and Refs.~\ocite{Stroem,Crepin12,GeissCrep14},
which had predicted absence and presence of backscattering, respectively,
from different approaches for the same, initially local fermionic model.

\section{Acknowledgements}

We thank J. C. Budich, F. Cr\'epin, F. Dolcini, M. S. Foster, I. V. Gornyi, N. Kainaris, P. Recher, T. L. Schmidt and N. Traverso Ziani
for helpful and inspiring discussions.
We appreciate financial support of the DFG (SFB 1170 and SPP 1666), the Helmholtz Foundation (VITI),
and the Elitenetzwerk Bayern (ENB graduate school on topological insulators).

\end{document}